\pdfoutput=1
\documentclass[12pt]{iopart}
\usepackage{graphicx}
\usepackage{hyperref}
\usepackage{subfig}
\usepackage{color}
\usepackage{amssymb}
\usepackage{adjustbox}

\usepackage[backend=bibtex,maxbibnames=10,style=phys]{biblatex}
\AtEveryBibitem{\clearfield{title}}
\bibliography{gwdetref,deeplearningref}

\definecolor{update}{RGB}{0, 158, 115}
\definecolor{check}{RGB}{213, 94, 0} 
\definecolor{intreview}{RGB}{0, 0, 0}
\definecolor{extreview}{RGB}{0, 0, 0}
\definecolor{extreview2}{RGB}{0, 0, 255}

\definecolor{todo}{RGB}{255, 0, 0}

\newcommand\ireviewme[1]{\textcolor{intreview}{#1}}
\newcommand\ereviewme[1]{\textcolor{extreview}{#1}}
\newcommand\ereviewmetwo[1]{\textcolor{extreview}{#1}}
\newcommand\ereviewmethree[1]{\textcolor{extreview}{#1}}

\begin{document}

\title[Image-based deep learning for classification of noise transients in gravitational-wave detectors]{Image-based deep learning for classification of noise transients in gravitational wave detectors}

\author{Massimiliano Razzano$^{1,3}$ and Elena Cuoco$^{2,3}$}

\address{Department of Physics, University of Pisa, Pisa I-56127, Italy}
\address{European Gravitational Observatory (EGO), I-56021 Cascina, Pisa, Italy}
\address{Istituto Nazionale di Fisica Nucleare, Sezione di Pisa, Pisa I-56127, Italy}

\ead{massimiliano.razzano@unipi.it}
\vspace{10pt}
\begin{indented}
\item[] \today
\end{indented}

\begin{abstract}
The detection of gravitational waves has inaugurated the era of gravitational astronomy and opened new avenues for the multimessenger study of cosmic sources. Thanks to their sensitivity, the Advanced LIGO and Advanced Virgo interferometers will probe a much larger volume of space and expand the capability of discovering new gravitational wave emitters. The characterization of these detectors is a primary task in order to recognize the main sources of noise and optimize the sensitivity of interferometers. Glitches are transient noise events that can impact the data quality of the interferometers and their classification is an important task for detector characterization. Deep learning techniques are a promising tool for the recognition and classification of glitches. We present a classification pipeline that exploits convolutional neural networks to classify glitches starting from their time-frequency evolution represented as images. We evaluated the classification accuracy on simulated glitches, showing that the proposed algorithm can automatically classify glitches on very fast timescales and with high accuracy, thus providing a promising tool for online detector characterization.

\end{abstract}

%
\vspace{2pc}

\section{Introduction}\label{sec:introduction}
The detection of gravitational waves by the Advanced Laser Interferometer Gravitational Observatory (LIGO) has started the era of gravitational wave astronomy and opened a new window on the Universe. \ireviewme{Among the events detected so far, \ereviewmetwo{most of them} are consistent with the coalescence of binary black holes (e.g. GW150914, GW151226, GW1701014, GW170814 \cite{2016gw150914,2016gw151226,2017gw170104,2017gw170814}), while GW170817 \cite{2017gw170817} is interpreted as the signal from coalescing neutron stars}.\\ 
The \ireviewme{coalescence} of compact binary systems is a primary source of gravitational waves. In particular, systems containing at least one neutron star are believed to be powerful \ireviewme{electromagnetic} emitters, thus being prime targets for multimessenger investigations. Core-collapse supernovae are also expected to emit bursts of gravitational waves \cite{2016gossansn}. Furthermore, continuous, periodic gravitational wave emission is expected from asymmetric neutron stars \cite{2017abbottknownpsr}, as well as a continuous gravitational-wave stochastic background.\\ 
Advanced LIGO \cite{2015aligo} and Advanced Virgo \cite{2015advirgo} are second-generation laser interferometers aimed at reaching a sensitivity $\sim$10 times better that of the previous LIGO and Virgo detectors. This increase in sensitivity will offer the possibility to explore a larger volume of space and increase the rate of detected signals. In order to reach the target sensitivity, both LIGO and Virgo detectors underwent significant upgrades on all major subsystems, including optics, suspensions, and seismic attenuation systems.\\ 
Advanced LIGO began operations \ireviewme{in 2015} and has completed two observing runs so far, called O1 (12 September 2015 - 19 January 2016) and O2 (30 November 2016 -- 25 August 2017). The construction of Advanced Virgo has been completed at the end of 2016 and the interferometer has joined O2 on 1 August 2017\cite{2016obsscenario}.\\
The detection of gravitational wave transients is based on two different strategies implemented by the LIGO and Virgo Collaborations with dedicated pipelines. When the gravitational waveform is known with a good accuracy, as for a compact binary coalescence, the optimal detection is achieved with  matched filtering with template banks of expected waveforms (e.g. \cite{2016pycbc,2016mbta}). When the shape of  gravitational waveform is unknown, excess-power strategies has been developed, such as coherent network algorithms \cite{2008cwb}.\\
All search methods are limited by the detector noise, which usually is non stationary and frequency-dependent. At low frequencies ($<$10 Hz) the main contribution to the noise is due to seismic ground motion and to the gradient of the local gravitational field. At higher frequencies the thermal noise of the subsystems, e.g. the suspensions of the mirrors, plays a dominant role. Above $\sim$200 Hz the most important limitation to the detector sensitivity is the shot noise of the laser light circulating in the detector cavities \cite{2011cella}.\\
The noise in the interferometers \ireviewme{has also} transient components produced by interactions among detector subsystems or with the surrounding environment. These transient noise signals, called ``glitches'' have an impact on the  \ereviewme{data quality} of the interferometers and in some cases could mimic the time-frequency behavior of astrophysical signals. However, real gravitational wave transient events produce consistent waveforms that are delayed due to the propagation between detectors, and so they can be discriminated from noise by comparing data in independent detectors \cite{2010s6trans,2012virgotrans}.  Near-simultaneous glitches can still mimic the signature of a gravitational wave signals, therefore it is also important to look for noise transients in auxiliary channels of each detector, that are not sensitive to gravitational waves but can provide useful information on the status of the environment and of the interferometer.\\
The identification and characterization of these glitches is important in order to investigate the impact of transient noise on the detectors. It is very useful to produce a classification of glitches according to their time and frequency evolution, in order to group them in families of similar \ireviewme{morphology}.\\ 
Once a glitch family has been identified, it is possible to perform further investigations to establish its origin and prepare custom data quality flags in order to reduce their impact on the detector performance \cite{2010s6trans}.\\ 
Manual inspection of the glitch parameters is time consuming and can be subject to errors, therefore is not the optimal strategy in the era of advanced interferometers, since the increased complexity and sensitivity of the detectors will lead to large quantities of glitches of various \ireviewme{morphologies}.\\
Different methods have been developed to automatically classify glitches in advanced gravitational wave detectors. \ereviewme{For instance, in \cite{2015powellclass}, three algorithms were tested on Advanced LIGO Engineering Run data : 1) PCAT, based on Principal Component Analysis \cite{2001calderpca}, 2) LALInferenceBurst, based on Bayesian parameter estimation \cite{2015essicklib}, and 3) WDF-ML, that couples a Wavelet Detection Filter \cite{2005acernesenap} with an unsupervised machine learning clustering based on Gaussian Mixture Model algorithm \cite{2011pedregosascikit}. In \cite{2017mukunddb}, a supervised Bayesian classifier called Difference Boosting Neural Network (DBNN) has been tested on LIGO S6 data, also showing very good performance.}\\
Among the possible approaches to the automatic analysis and classification of glitches, machine learning looks very promising. Machine learning is a field of computer science based on algorithms that can learn from data and make predictions on it \cite{1959Samuel}. This approach is extensively used in many applications of Big Data analysis, including marketing, speech recognition and computer vision, and is becoming very popular in many field of scientific research, \ereviewme{including the study of electron structure \cite{2017brockherde,2017li}, quantum many-body systems \cite{2017shutt}, molecular models \cite{2016Kearnes,2017faber}, and particle physics \cite{2017alves}}.\\
Machine learning algorithms can be divided in \ereviewme{two main} categories: supervised ones, which perform classification or regression tasks by learning from samples of labeled data, and unsupervised ones, which learn relationships automatically from non-labeled data.\\
Deep learning is an area of machine learning that combines the architecture of the artificial neural networks with the power of the machine learning algorithms (see, e.g. \cite{2015dlover}). By stacking many layers of artificial neurons in a neural network architecture, it is possible to build very efficient tools that perform classification or regression tasks.\\
Deep learning algorithms have been suggested for applications in gravitational wave physics \cite{2017george}, highlighting the potential of convolutional neural networks (CNNs).
In particular, the time-frequency behavior of glitch signals can be converted to images and given as input to CNNs, which show very good performance in image analysis and recognition tasks \cite{2016alexnet}.\\ 
\ereviewme{Various baseline methods exist to classify images, such as minimum distance, maximum likelihood, Support Vector Machines, Random Forest \cite{2017george}, or fully connected neural networks. 
However, we decided to explore the potential of deep learning because of the complexity of gravitational wave data, as done in previous works (e.g. \cite{2015powellclass,2017george}). In particular, transient noise glitches often originate from complex nonlinear couplings between the channels of the detector, thus resulting in very complex time-frequency morphologies. This is supported by the analysis of real data, where glitches appear to come in many different classes.
Furthermore, the size of the time-frequency images that we are using to classify glitches is too large for a simpler method (e.g. fully connected networks), given the constraints on our hardware.
The exquisite capability of deep networks to extract features from large images with minimal setup and parameter tuning, and their power in multiclass problem, makes them an optimal algorithm for our purpose.
Training a deep network could take time, but this is not an issue for our analysis. In fact, to classify a set of glitches without any requirement on speed, the training can be performed once. In case the classification algorithm is running in low-latency mode on new coming data, and thus speed is fundamental, the training can be performed periodically on a dedicated machine, and the resulting model can be uploaded to a second machine that is dedicated only to the classification.\\}
In this \ireviewme{paper} we introduce a new tool for the classification of transient noise in gravitational wave interferometers, that exploits the power of image processing with CNNs. The structure of the \ireviewme{paper} is the following: in section 2 we introduce the deep learning algorithms, focusing on the CNNs and their utility in astrophysics. In section 3 we describe the proposed analysis pipeline and algorithms, while section \ref{sec:simulations} is focused on the simulated transients prepared to test the method. In section \ref{sec:results} we report the summary of the performance, that are further discussed in section \ref{sec:discussion}. In section \ref{sec:conclusions} we highlight the possible extensions and prospects of the suggested method.
\section{Deep learning for image recognition}
Deep learning has found many applications in the analysis of Big Data, in particular on complex tasks such as computer vision \cite{2017Ioannidoucvsurvey}, image processing \cite{2016Druzhkov}, speech recognition \cite{2017zhangsppech} and natural language processing \cite{2011Collobertnlp}. Furthermore, deep learning algorithms for image analysis and recognition have been successfully tested and implemented in many fields of astrophysics, including galaxy classification \cite{2017lukicdlgalaxy}, cosmology \cite{2017schmelzecosmo}, and asteroseismology \cite{2017stello}.\\
The base element of a neural network is the artificial neuron, introduced in 1943 by McCulloch and Pitts as a result of their efforts to understand and design artificial networks \cite{1943McCulloch}. A later work introduced the {\it perceptron} rule \cite{1958RosenblattPerceptron}, an algorithm that automatically learns the optimal weights coefficients w$_{i}$ that are multiplied by the components of an input vector $\vec{\textbf{x}}$ in order to decide whether a neuron will fire or not.\\ 
In fact, each perceptron computes the {\it activation function} $\phi(z)$, where $z$ is the so-called {\it net input} defined as:
\begin{equation}
z = \vec{\textbf{w}} \cdot \vec{\textbf{x}}+b = \sum\limits_{i=1}^n w_{i}x_{i}+b
\end{equation}
In the context of a single-neuron network, $\vec{\textbf{x}}$ represents the vector containing the properties, or {\it features}, of a data sample. For image analysis, the features could be the intensity values of each pixel. In more complex networks, the vector $\vec{\textbf{x}}$ is the outputs of the other perceptrons. The coefficient $b$ is the {\it bias} term and is introduced to shift the decision boundary, which can be useful to improve the accuracy of the algorithm. In the basic implementation of the perceptron, the activation function is the Heaviside step function $\theta(z)$ and can be used to implement a classifier among two classes corresponding to the two output values of that function.\\
The interest in neural networks was boosted by the development of the {\it backpropagation} technique for efficient training of neural networks \cite{1986rumelhartbackp}. The introduction of the {\it multilayer perceptron}, an artificial network made of stacked layers of neurons, overcame the limitations of the perceptron. These architectures are also called {\it feedforward neural networks} because they process the input forward \ereviewme{to the output without forming any internal cycle.}\\
Furthermore, the universal approximation theorem \cite{1989Hornikuat} states that a 3-layer feedforward network with one hidden layer and finite number of neurons can approximate any continuous function. Therefore, deep neural networks can be used to model and approximate complex functions, whose output could be discrete or continuous, for classification or regression tasks respectively.\\
Convolutional Neural Networks (CNNs) are a subset of feedforward deep neural networks developed to improve the performance of deep learning on image processing \cite{1998lecuncnn}. The key element in CNNs are {\it convolutional} layers, where neurons are optimized to capture specific details in an image. In order to achieve this, the neurons are sharing weights within a kernel filter. During the training, the weights are computed by sliding the filter across the neurons in the layer, resulting in a kernel convolution. The size of the kernel is related to the scale of the details to be looked for in the image. Sharing the weights greatly reduces the number of neurons required to analyze an image, with a net gain in computational resources with respect to the standard, fully-connected, deep networks.
\ereviewme{Furthermore, sharing weight also introduces a regularization, thus preventing the algorithm from potential overfitting.}\\
Each convolutional layer is coupled with a {\it pooling} layer that acts as a subsampling operator by reducing the size of the output of a convolutional layer.\\ 
In order to reduce the problem of overfitting, CNNs also feature {\it dropout} layers, that randomly ignore a set of neurons during training \cite{2012hinton}. The dropout is a regularization strategy widely used and has the overall effect of making the CNN less sensitive to the specific weights of neurons.\\
Once a CNN has been trained on a set of images, it is capable of \ireviewme{performing} classification or other tasks in very short time, and is therefore very well suited for tasks that require low-latency processing. Thanks to their flexibility and capability of processing large images, CNNs have been widely implemented in various fields of Medical Physics \cite{2016rao}, and it is now being tested in projects involving image analysis in astronomy \cite{2016flamary}. The accuracy and speed of CNNs are also very important for  \ereviewme{tasks} requiring low latency, and they have been proposed also for the automatic search for transient phenomena in gravitational wave data \cite{2017george,2017georgetransfer}.
\section{Description of the pipeline}
We implemented a CNN-based classification pipeline for the automatic classification of glitches and other noise transients in advanced gravitational wave detectors. The input of the pipeline are images representing the time-frequency evolution of the glitches
For instance, this can be derived from the signal in 
h(t), the channel representing the output of the interferometers after proper vetoes and calibrations are applied. For a transient signal of astrophysical origin, h(t) measures the gravitational wave strain induced by the passage of gravitational waves, and can be contaminated by glitches.
\subsection{Whitening procedure}\label{sec:whitening}
Before applying any detection and classification pipeline, the data should be conditioned. The usual procedure is the whitening one \cite{2001bcuoco,2004cuoco}, which can remove the contribution of Gaussian colored  detector noise. Many pipelines use whitening in frequency domain, while the whitening here applied is based on time-domain procedure, using an AutoRegressive (AR) fit to the data.   
 
An Auto-Regressive process $x[n]$ of order $P$ with parameter $a_k$, from here after $AR(P)$, is characterized by the relation
\begin{equation}
\label{eq:AR}
x[n]=\sum _{k=1}^{P}a_{k}x[n-k]+\sigma w[n]\, ,
\end{equation}
being $w[n]$ a white Normal process.

The problem of determining the AR parameters is the same of that of finding
the optimal ``weights vector'' ${\mathbf w}=w_k$, for $k=1,...P$  for the
problem of linear prediction \cite{1988Kay_book}. 
In the linear prediction we would predict the sample $ x[n] $ using the $ P $
previous observed data ${\mathbf x}[n]=\{x[n-1],x[n-2]\ldots x[n-P]\} $
building the estimate as a transversal filter:
\begin{equation}
\hat{x}[n]=\sum _{k=1}^{P}w_{k}x[n-k]\, .
\end{equation}
 
We choose the coefficients of the linear predictor by minimizing  a cost
function that is the mean squares
error $ \epsilon ={\mathcal{E}}[e[n]^{2}] $ (${\mathcal{E}}$ is the operator of average on the ensemble), being 
\begin{equation}
\label{eq:error}
e[n]=x[n]-\hat{x}[n]
\end{equation}
the error we make in this prediction and obtaining the so called Normal or
Wiener-Hopf  equations
\begin{equation}
\label{eq:Normal}
\epsilon _{min}=r_{xx}[0]-\sum _{k=1}^{P}w_{k}r_{xx}[-k]\, ,
\end{equation}
 which are identical to the Yule--Walker equations \cite{1988Kay_book} used to estimated the parameters $a_k$ from autocorrelation function with $w_{k}  = -a_{k}$ and $\epsilon _{min}  =  \sigma ^{2}$

This relationship between AR model and linear prediction assures us to obtain
a filter which is stable and causal \cite{1988Kay_book}, so we can use the $AR$ model to reproduce stable processes in time-domain. It is this relation between
AR process and linear predictor that becomes important in the building of whitening
filter.
\ireviewme{The tight relation between the AR filter and the whitening filter is clear in the figure~\ref{fig:ar-predic}. The figure describes how an AR process colors
a white process at the input of the filter if you look at the picture from left
to right. If you read the picture from right to left you see a colored process
at the input that pass through the AR inverse filter coming out as a white process.
\begin{figure}[ht]
\centering 
\subfloat[]{\includegraphics[width=0.5\textwidth3]{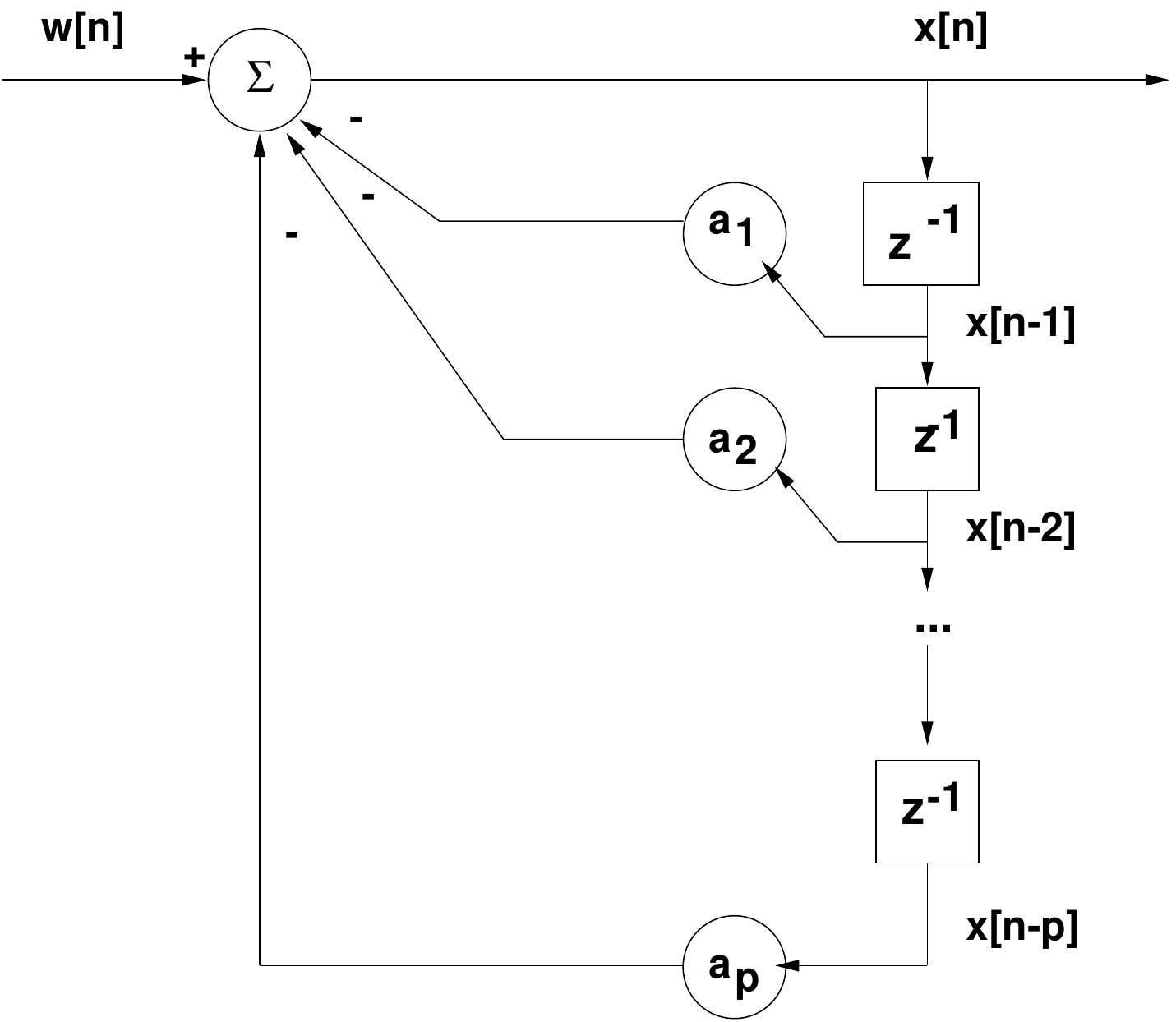}}
\caption{\label{fig:ar-predic} Whitening filter and AR filter.}
\end{figure}
}
When we find the $ P $ parameters that fit a PSD of a noise process, what
we are doing is to find the optimal vector of weights that let us reproduce the process
at the time $ n $ knowing the process at the $ P $ previous time. All
the methods that involve this estimation try to make the error signal (see
equation (\ref{eq:error}) ) a white process in such a way to throw out all the correlation
between the data (which we use for the estimation of the parameters).
Suppose to have a sequence $x[n]$ of data characterized
by an autocorrelation $r_{xx}[n]$ which is not a delta function,
and that you need to remove all the correlation, making it a white
process. The idea is to model $x[n]$ as an AR process,
find the AR parameters and use them to whiten the process.\\ 
Using a lattice structure \cite{2001bcuoco} we can implement the whitening filter in time domain. This is a procedure which is used for other pipeline \cite{2007acernesegrb,2015powellclass} and which can be implemented also in adaptive way \cite{2001acuoco} taking care of non stationary noise. 
\subsection{Image preparation}\label{sec:preparation}
If the images of the glitches are not available, we introduce an image preparation stage, that converts the whitened time series to images showing the time-frequency evolution of the glitches. 
In order to test the pipeline, \ereviewme{in this paper we have}  built images using simulated time series that cover 2 seconds around each glitch. A complementary, multi-view architecture combining different timescales has also been proposed \cite{2017gravityspy}. The images are built from the Fourier spectrograms of the time series. Other algorithms can be used to produce the images, such as the constant Q transform or Wavelet transforms. The spectrograms are then converted to 32-bit gray scale images, by rescaling the spectrogram values in the range 0-2$^{32}$.\\
The images can be further processed, if needed, in order to reduce the residual noise features and enhance the details. 
\ireviewme{For instance, in order to further enhance the contrast in the images we can apply, where needed, a normalization in the image histogram that ignores the pixel values outside the 2\% and the 98\% percentiles.}
\subsection{Detection and Classification}\label{sec:detection}
The main goal of this algorithm is the classification, however CNNs can be also used to detect glitches in the data stream. The detection of a \ereviewme{glitch} signal can be seen as an {\it anomaly detection} problem, that can be tackled with supervised or unsupervised algorithms. In our framework of supervised CNNs, a simple way to implement the anomaly detection is using \ereviewme{a ``signal vs noise'' binary classification approach\footnote{From now on, by \emph{signal} we mean a transient glitch signal, and for \emph{noise} we mean the stationary, non-transient, detector noise}. To this purpose, the images containing glitches can be grouped in a SIGNAL class, and a set of time-frequency images of data stream with no glitches can be used to produce a set of images for a NOISE class, since they contain only the non transient component of the detector noise}.
This binary classification problems work well for standard machine learning algorithms, such as Support Vector Machines or Random Forest.\\
\ereviewme{Since CNNs work smoothly on predicting many classes, we implemented in our pipeline a multi-class classification. Given a number of glitch classes $N_{c}$, we have produced a set of glitch-free images to populate a NOISE class, so that the pipeline can perform a $N_{c}$+1 classification of glitches or noise. For each glitch image, the pipeline computes the probability of belonging to one of the $N_{c}$ glitch class or to the NOISE class\footnote{Even if the presence of a glitch in the data stream has been already assessed by another external detection algorithm, the presence of a NOISE class is useful to cross check the robustness of the detection.}.}
For the training phase we split our glitch set in training\ereviewme{, validation, and test set with the ratio 70:15:15}.\\
In a realistic situation, it can happen that the number of images is different for each class \cite{2017gravityspy}. To fix this class imbalance problem, we implemented a balancing step based on image augmentation \ireviewme{procedures \cite{2016imageaugmentation}}. \ereviewme{Given the appearance of the time-frequency images of glitches, we perform the augmentation by simply shifting the images by a small amount, typically of $\sim$10\% of the size of the image. We have found that rotating and flipping the images is not very important on glitch images, that show always the same orientation and direction in time.}\\
Once the training phase has been completed, the model and the weights computed by the CNN are saved on disk. Using this model, the pipeline can perform classification starting from the time series or the image of the new glitches. The output of this step is the class of the glitch, with the associated probability of belonging to that class.
\subsection{Implementation of the architecture}\label{sec:implementaion}
The pipeline has been developed using the Python scripting language, and we used high-performance numerical libraries such as NumPy and SciPy \cite{numpyref,scipyref}. The CNN architecture has been built using the Python-based libraries Theano \cite{2016theano} and Keras \cite{2015keras}, which provide high performance and easy implementation of prototypes. The code has been developed and tested on a GPU NVIDIA GeForce GTX 780 running CUDA 8.0 and and the accelerated libraries cuDNN developed for deep learning applications  \cite{2014cudnn}.\\
\begin{figure}[ht]
\centering 
\subfloat{\includegraphics[width=0.45\textwidth]{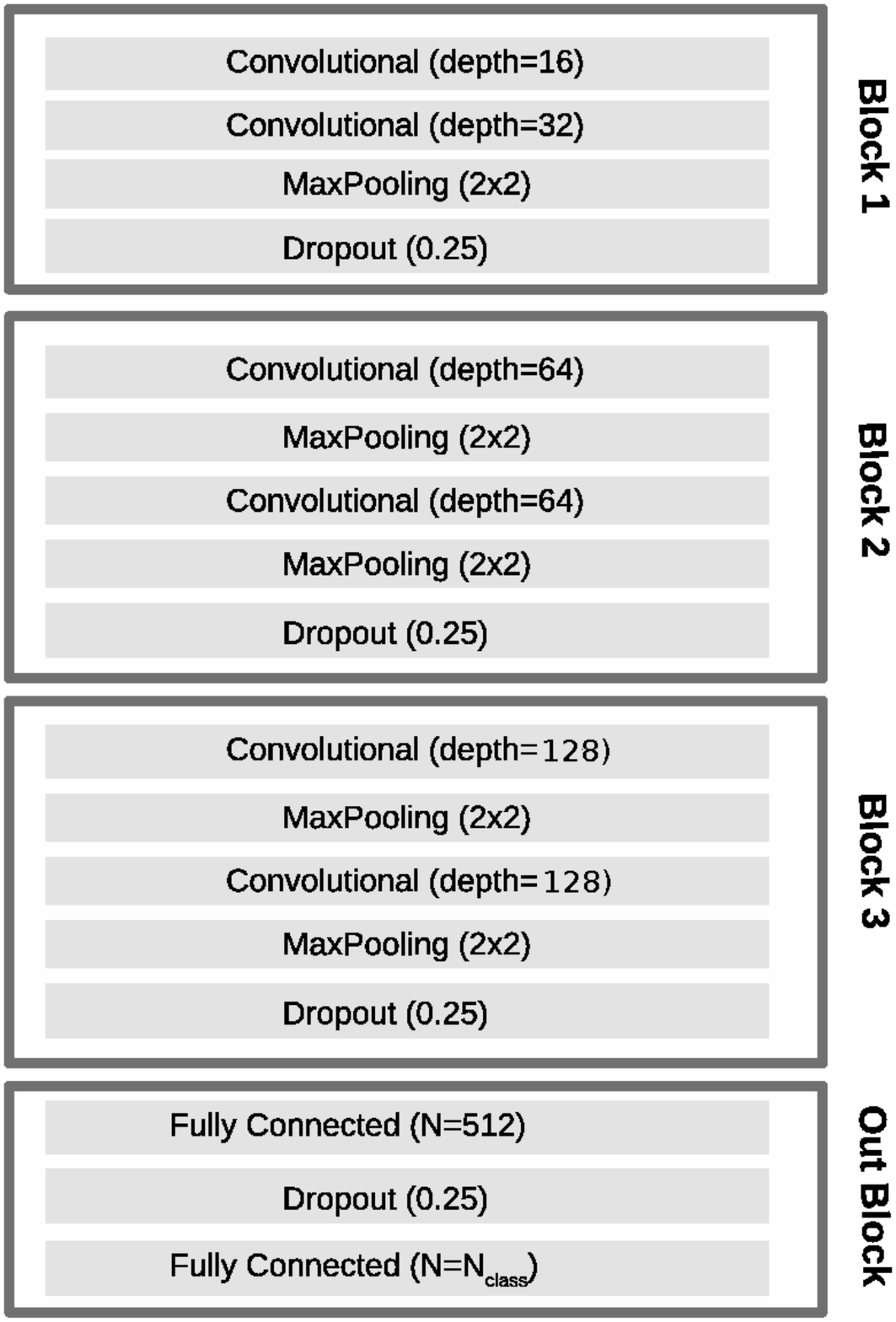}}
\caption{\label{fig:architecture} Diagram showing the structure of the CNN used in this work (For more detail see section \ref{sec:implementaion}). \ereviewme{The size of the convolutional kernels is always 3x3. Here the output size is the number of classes $N_{c}$, but becomes $N_{c}$+1 if the NOISE class is used (See text for more details)}}
\end{figure}
\ereviewmetwo{We have tested difference CNN configurations and we selected an architecture based on four main blocks.} The first block is made of a 16 and a 32-deep layer, while the second and the third block contains 2 layers with depth of 64 and 128 respectively. Each convolutional kernel has a 3 $\times$ 3 and a \emph{ReLU} activation function.
Each block contains Maximum Pool layers \ereviewme{(Keras MaxPooling2D)} and is followed by a dropout layer \ereviewme{(Keras Dropout)}. The last block is made of a fully connected layer \ereviewme{(Keras Dense)} with 512 neurons and by an output layer with a size equal to the number of classes $N_{c}$ and with a {\it SoftMax} activation function. If the NOISE class is added, the output size is increased by one unit.
Figure \ref{fig:architecture} shows the blocks in the CNN used for the pipeline.\\
\ereviewmetwo{In order to show the classification capabilities of this pipeline, it is important to compare its performance with that of simpler architectures and use a common dataset for the comparison. The datasets used in previous works (e.g. those cited in section \ref{sec:introduction}) are in some cases too small for a deep learning algorithms or not available. Therefore, in order to assess the performance of our algorithm in this paper we have performed a comparison using the dataset of simulated glitches described in next section. The detail and results of this comparison will be presented in section \ref{sec:results}}
\section{Glitch simulations}\label{sec:simulations}
In order to test the classification algorithm, we prepared simulated a set of time series that include glitches with different shapes and intensities. This approach, also adopted in previous works \cite{2015powellclass,2017mukunddb}, allows to evaluate classification performance depending on the parameters of the simulated glitches.\\
The simulated glitches have been added to a Gaussian noise generated from the sensitivity curve of real interferometers. Here we used the sensitivity of LIGO Hanford detector (H1) available online\footnote{Data publicly available at the LIGO Open Science Center \url{https://losc.ligo.org/s/events/GW150914/P150914/fig1-observed-H.txt}}.
\subsection{Glitch families}\label{sec:glitchfam}
\begin{figure}[ht]
\centering 
\subfloat{\includegraphics[width=0.45\textwidth]{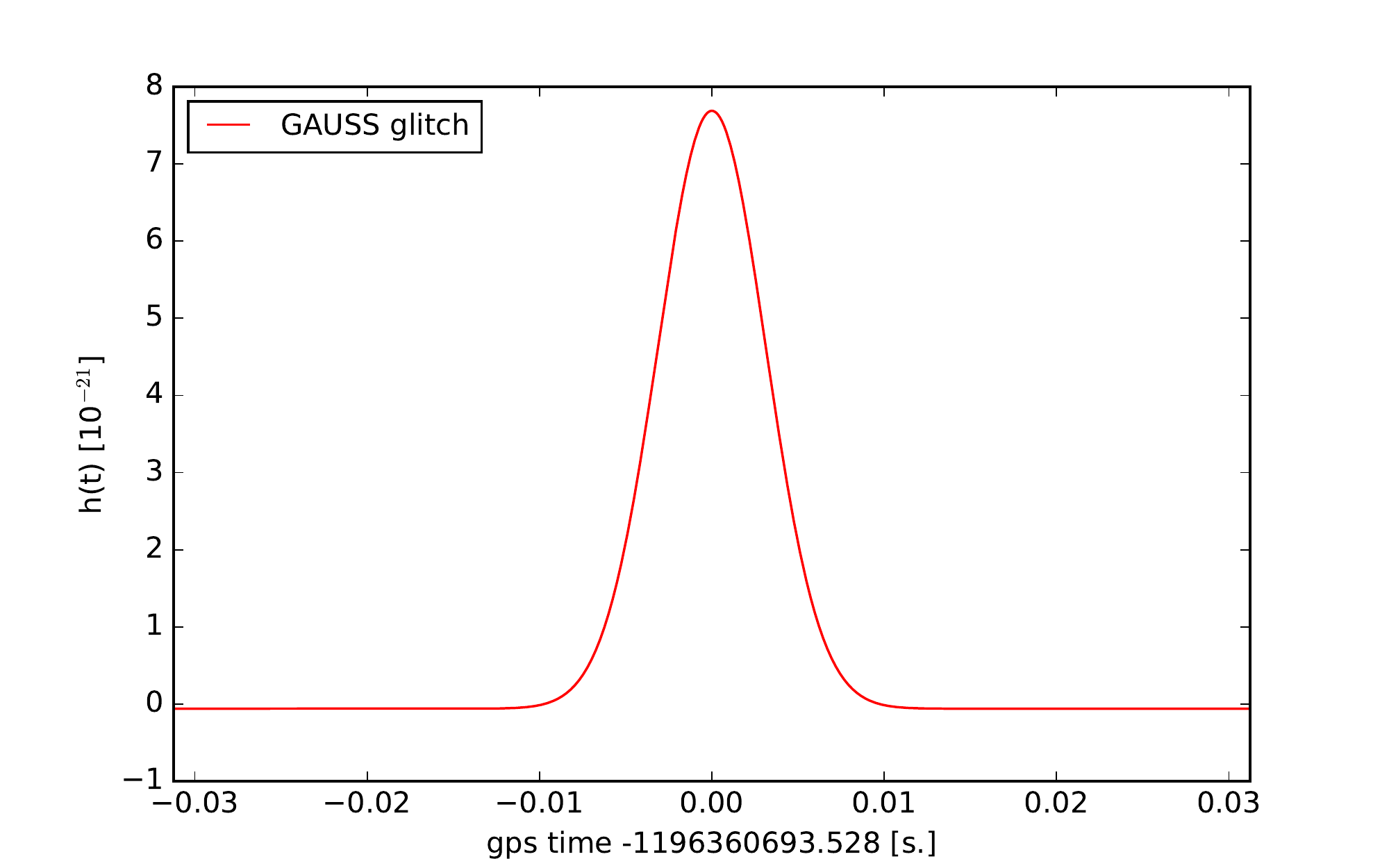}}
\subfloat{\includegraphics[width=0.45\textwidth]{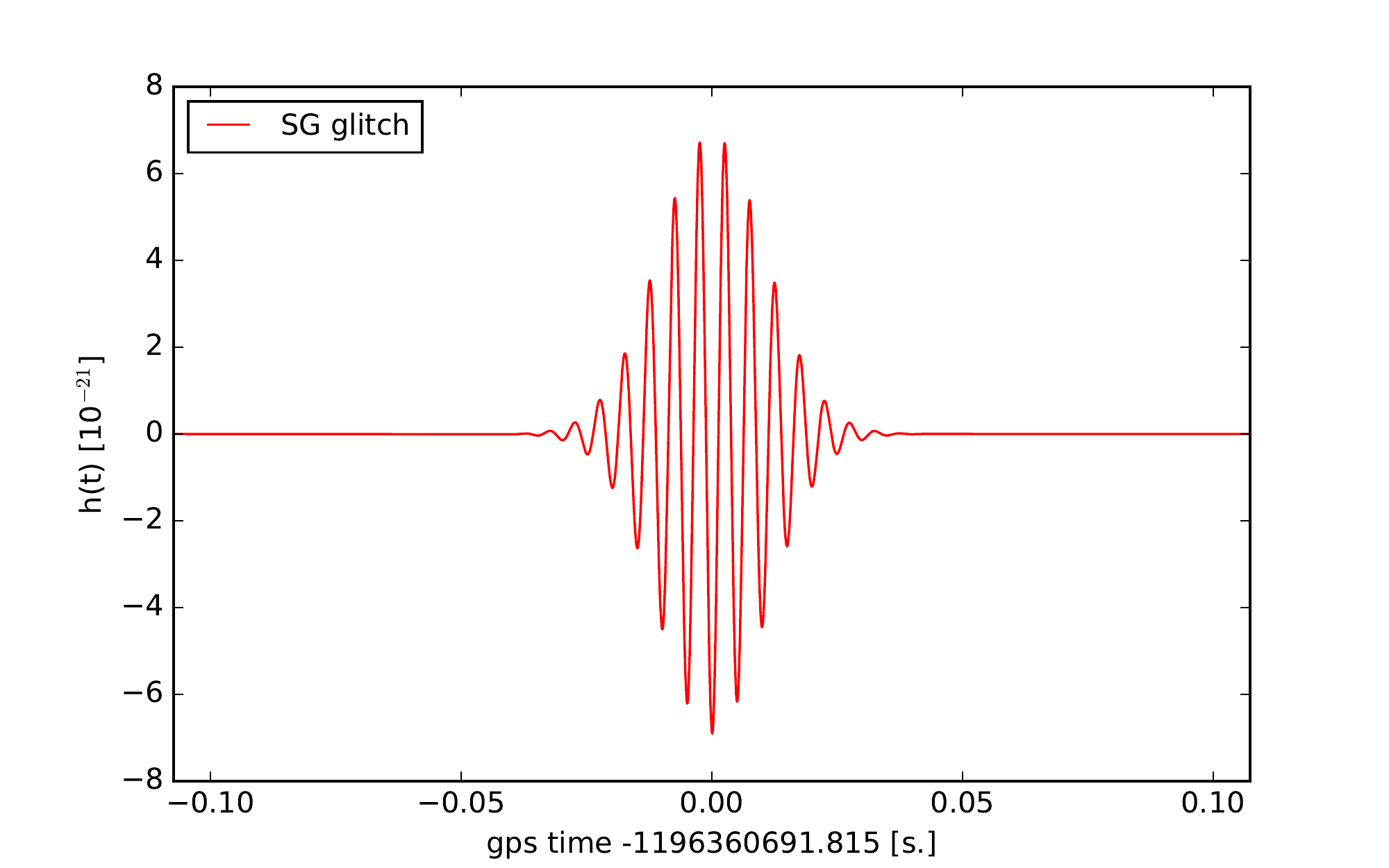}}
\hspace{0mm}
\subfloat{\includegraphics[width=0.45\textwidth]{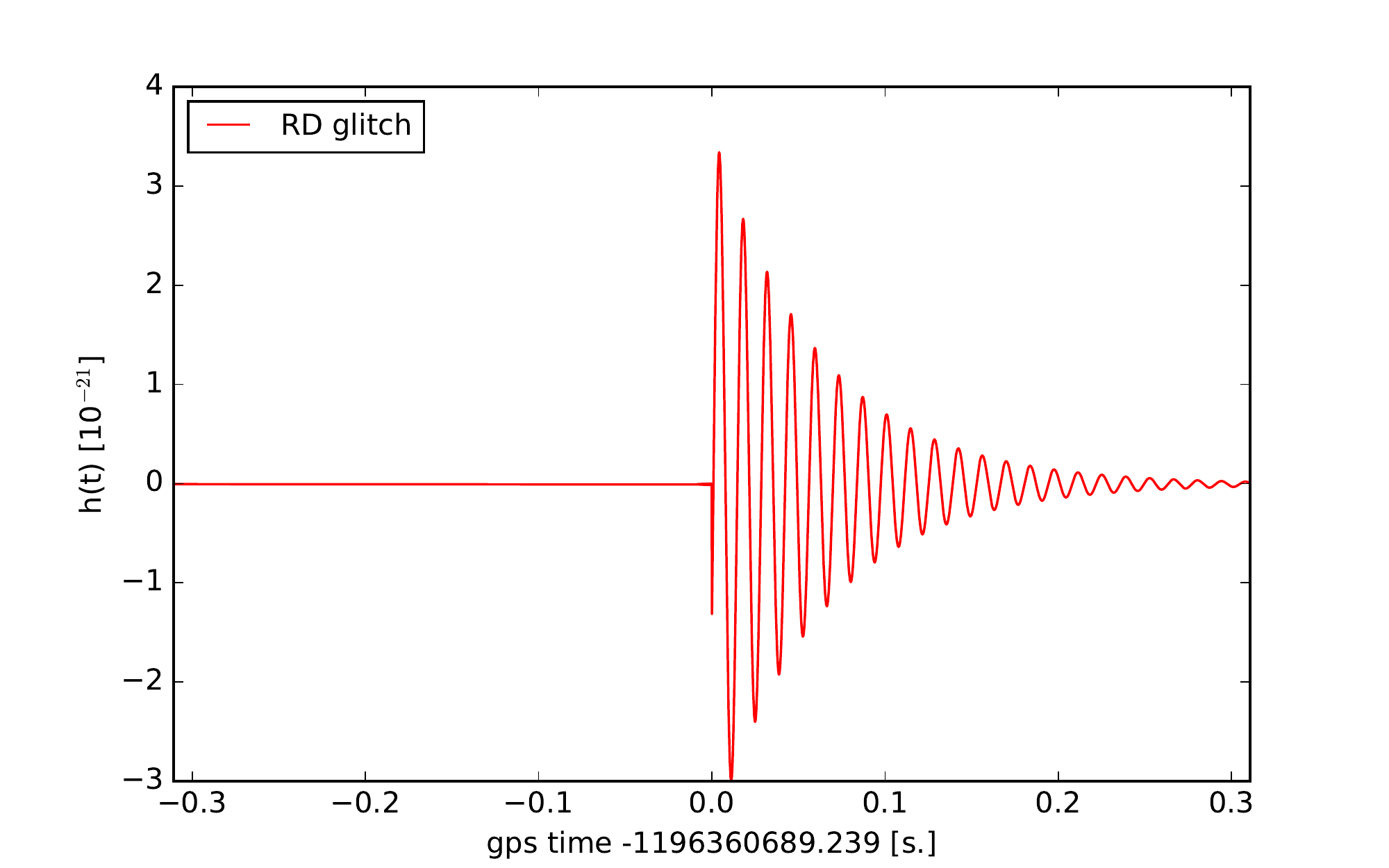}}
\subfloat{\includegraphics[width=0.45\textwidth]{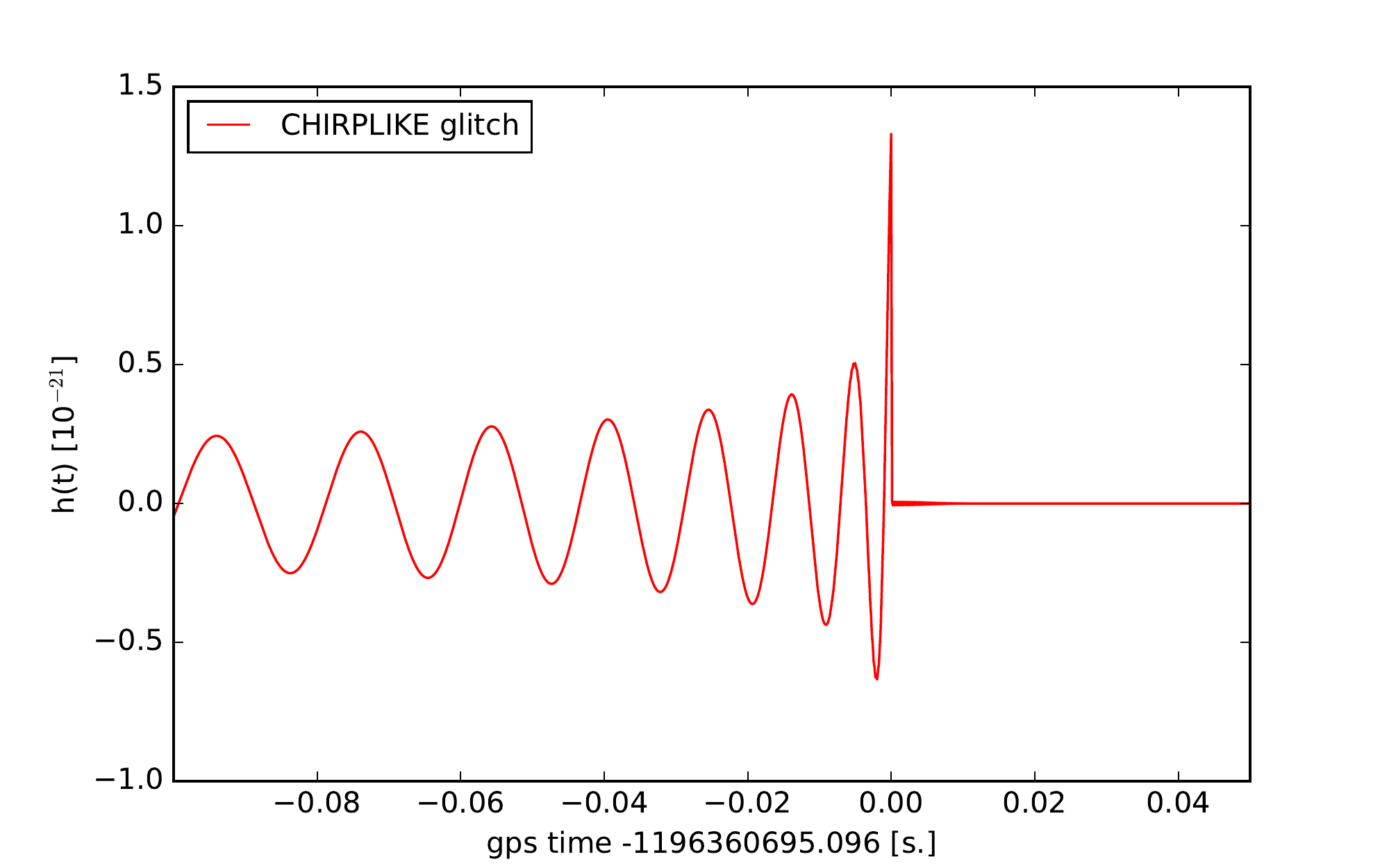}}
\hspace{0mm}
\subfloat{\includegraphics[width=0.45\textwidth]{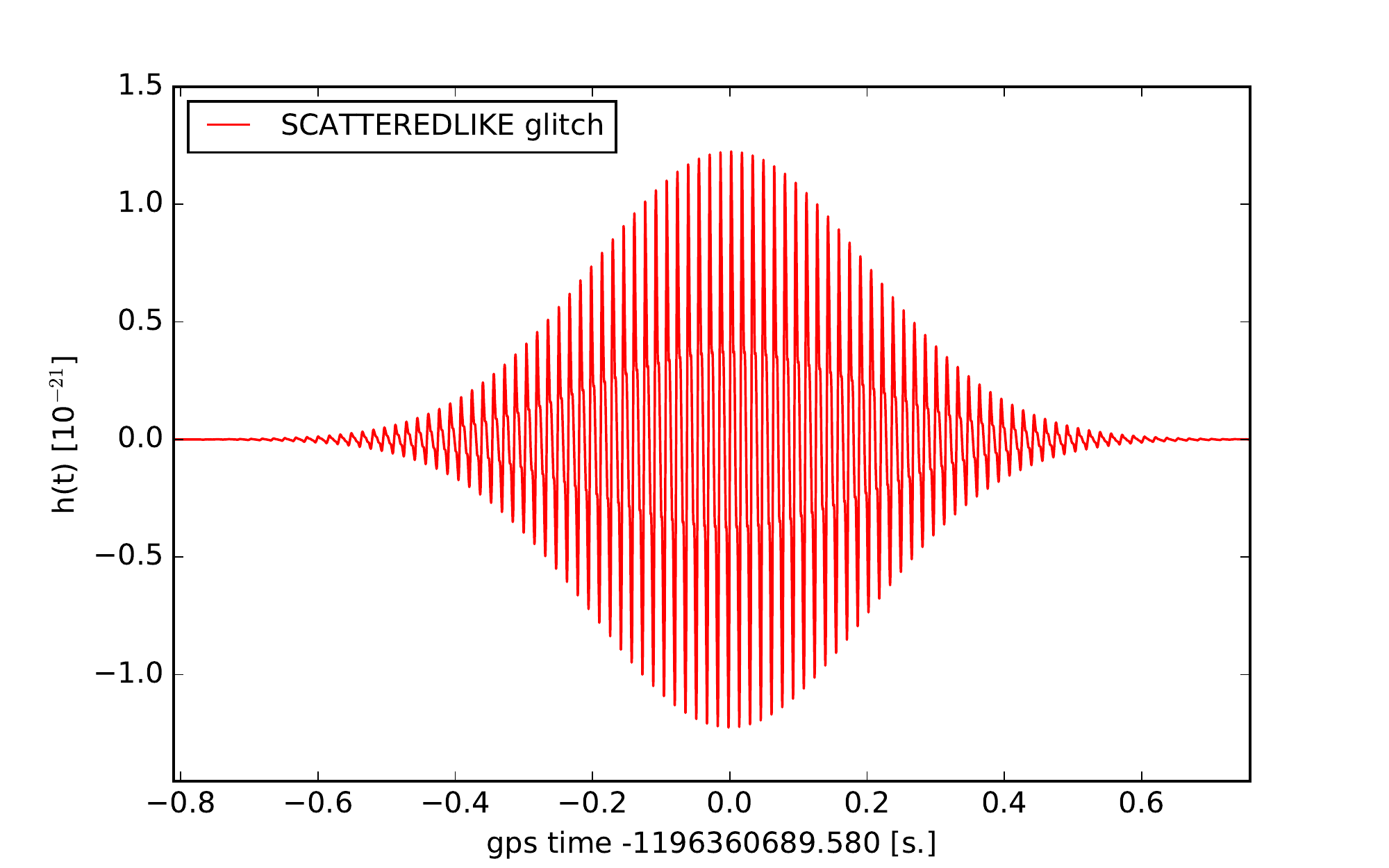}}
\subfloat{\includegraphics[width=0.45\textwidth]{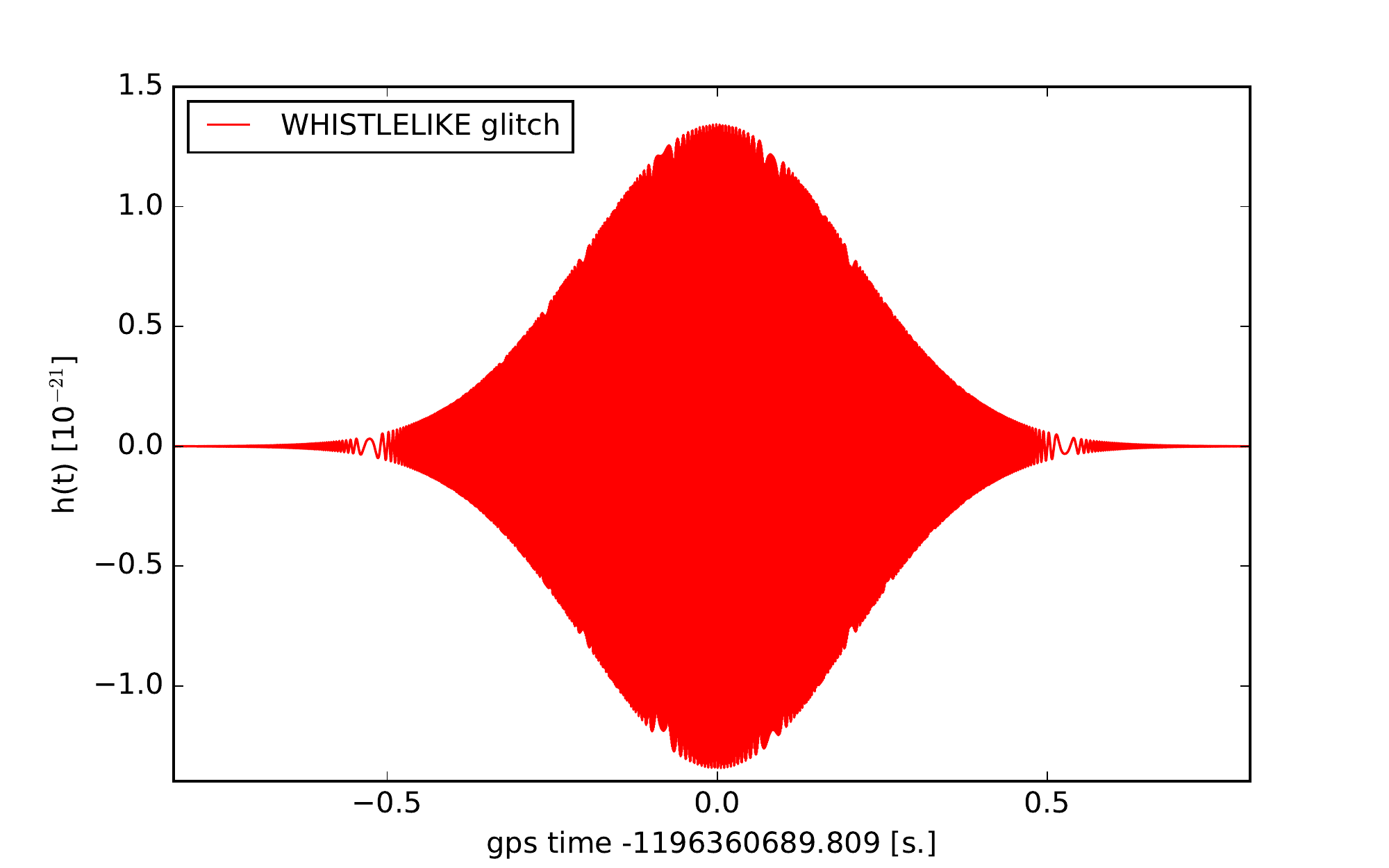}}
\caption{\label{fig:glitchgallery}Gallery of simulated glitches used for this study. We show only high-amplitude glitches without the contribution of the Gaussian noise in order to highlight the temporal evolution of the transients. The detailed description of each glitch family is in section \ref{sec:glitchfam}}
\end{figure}

\ireviewme{We have simulated six families of signals, that approximate the time-frequency evolution of main noise transients observed in the real interferometers \cite{2010s6trans}.\\ 
Each glitch is described by a time series h(t). We also included a family of chirp-like transients, in order  to show the potential of the CNNs in distinguishing transients of astrophysical origin and of known waveform, such as those produced by coalescing binaries. Therefore, we modeled these transients using 
a h$_{+}$(t) and h$_{\times}$(t) component, that are combined with the detector response to get the input signal h(t) = F$_{+}(\theta,\phi$) h$_{+}$(t) + F$_{\times}(\theta,\phi$) h$_{\times}$(t), where F$_{+}(\theta,\phi$) and F$_{\times}(\theta,\phi$) are the antenna factors of the interferometer. In this case,  $\theta$ and $\phi$ represent a random direction in the sky. Since chirp-like glitches have been observed in real data \cite{2010s6trans}, these simulations also help to show the capability of recognizing this glitch morphology. We then scaled the amplitude h$_{0}$ of each simulated glitch by its root sum squared amplitude h$_{rss}$ = $\int_{-\infty}^{\infty} (|h_{+}(t)|^{2}+|h_{\times}(t)|^{2})$dt \cite{2009abbottbursts}.\\ 
The glitch families are described below in more detail, while} figure \ref{fig:glitchgallery} shows a sample gallery of glitches belonging to these families, that we describe here:
\subsubsection[g]{Gaussian (GAUSS)}
These glitches are described by a simple Gaussian waveform and model the short spikes observed in previous works of detector characterization \cite{2010s6trans}. They have a characteristic width $\tau$, a peak time $t_{0}$, and can be modeled as:
\begin{equation}
\ireviewme{h(t)} = h_{0}\exp{\bigg(-\frac{(t-t_{0})^{2}}{2\tau^{2}}\bigg)}.
\end{equation}
\subsubsection[g]{Sine-Gaussian (SG)}
These glitches are described by a sinusoid whose amplitude is modulated by a Gaussian. They are centered at a frequency f$_{0}$ and their width $\tau$ is related to $f_{0}$ and to the quality factor $Q$ as $\tau$ = $\frac{Q}{\sqrt{2}\pi f_{0}}$.
\begin{equation}
\ireviewme{h(t)} = h_{0}\sin\bigg( 2\pi f_{0}(t-t_{0}) \bigg) \exp{\bigg(-\frac{(t-t_{0})^{2}}{2\tau^{2}}\bigg)}.
\end{equation}
\subsubsection[g]{Ringdown (RD)}
These signals are short in time and have limited bandwidth, and can be modeled by damped sinusoids with parameters defined as for SG glitches. For times t$>$t$_{0}$ RD can be written as:
\begin{equation}
\ireviewme{h(t)} = h_{0}\sin\big( 2\pi f_{0}(t-t_{0}) \big) \exp{\bigg(-\frac{(t-t_{0})^{2}}{2\tau}\bigg)}.
\end{equation}
\subsubsection[g]{Chirp-like (CL)}
\ireviewme{The chirping time evolution is very similar to the waveform produced by the coalescence of two compact objects such as neutron stars or black holes. Including this chirp-like family of glitches can show the potential of the method in discriminating noise from astrophysical signals, although the most robust way requires the comparison between the signals detected by different detectors \cite{2016gw150914}. This time evolution also reproduces the chirp-like observed in real data, that appears to be related to hardware injections.}\cite{2016zevingravityspy}. We modeled these glitches using the parameters of the coalescence of a compact systems, i.e. the masses m$_{1}$ and m$_{2}$ of the components and the time of the coalescence t$_{0}$ as the time of the simulated glitch. From these parameters we can compute the chirp mass $\mathcal{M}_{c}$ = $\frac{(m_{1}m_{2})^{3/5}}{(m_{1}+m_{2})^{1/5}}$ and the the time to coalescence $\tau_{c} = t - t_{0}$, useful to express the chirp-like transients as:
\begin{equation}
h_{+}(t) = h_{0}\tau_{c}^{-1/4}\cos\phi(t),
\end{equation}
\begin{equation}
h_{\times}(t) = h_{0}\tau_{c}^{-1/4}\sin\phi(t).
\end{equation}
where the phase $\phi(t)$ is computed as:
\begin{equation}
\phi (t) = -2 \bigg(\frac{5 G \mathcal{M}_{c}}{c^{3}}\bigg)^{-\frac{5}{8}}\tau_{c}^{\frac{5}{8}}+\phi_{0}.
\end{equation}
Without losing generality, for our simulations we fix  $\phi_{0}$ = 0.
\subsubsection[g]{Scattered light-like (SL)}
These simulations aim at reproducing the glitches associated to scattered light in the interferometers. In the time-frequency plane, these glitches appear at low frequencies as a series of arcs that can last up to $\sim$ 1 s \cite{1996vinetsl,2016zevingravityspy}. We modeled this long-lasting glitches with a modified quadratic chirp centered at $t_{0}$ and modulated with a Gaussian of characteristic width $\tau$. We simulate the glitch at a base frequency f$_{0}$ and its second and third harmonic. 
\begin{equation}
\ireviewme{h(t)} = h_{0}\sin(\phi_{SL})\exp{\bigg(-\frac{(t-t_{0})^{2}}{2\tau}\bigg)},
\end{equation}
Where the input $\phi_{SL}$ is defined as:
\begin{equation}
\phi_{SL} = 2 \pi f_{0} (t-t_{0})[1-K(t-t_{0})^{2}].
\end{equation}
We found that a value of $K$ = 0.5 is good in reproducing the curvature of the scattered light glitches.
\subsubsection[g]{Whistle-like (WL)}
Whistles glitches have been observed in Advanced LIGO \cite{2016zevingravityspy} and are related to MHz radio signals beating with the Voltage Controlled Oscillators. They occur at high frequencies and can extend for $\sim$ 0.5 s. We modeled them using the same functional form as scattered light, but fixing the curvature constant to a larger value related to the glitch width $K$ = 3$\tau$, and removing its harmonics.
\subsection{Simulated glitch sets}
\begin{figure}[ht]
\centering
\includegraphics[width=0.9\textwidth]{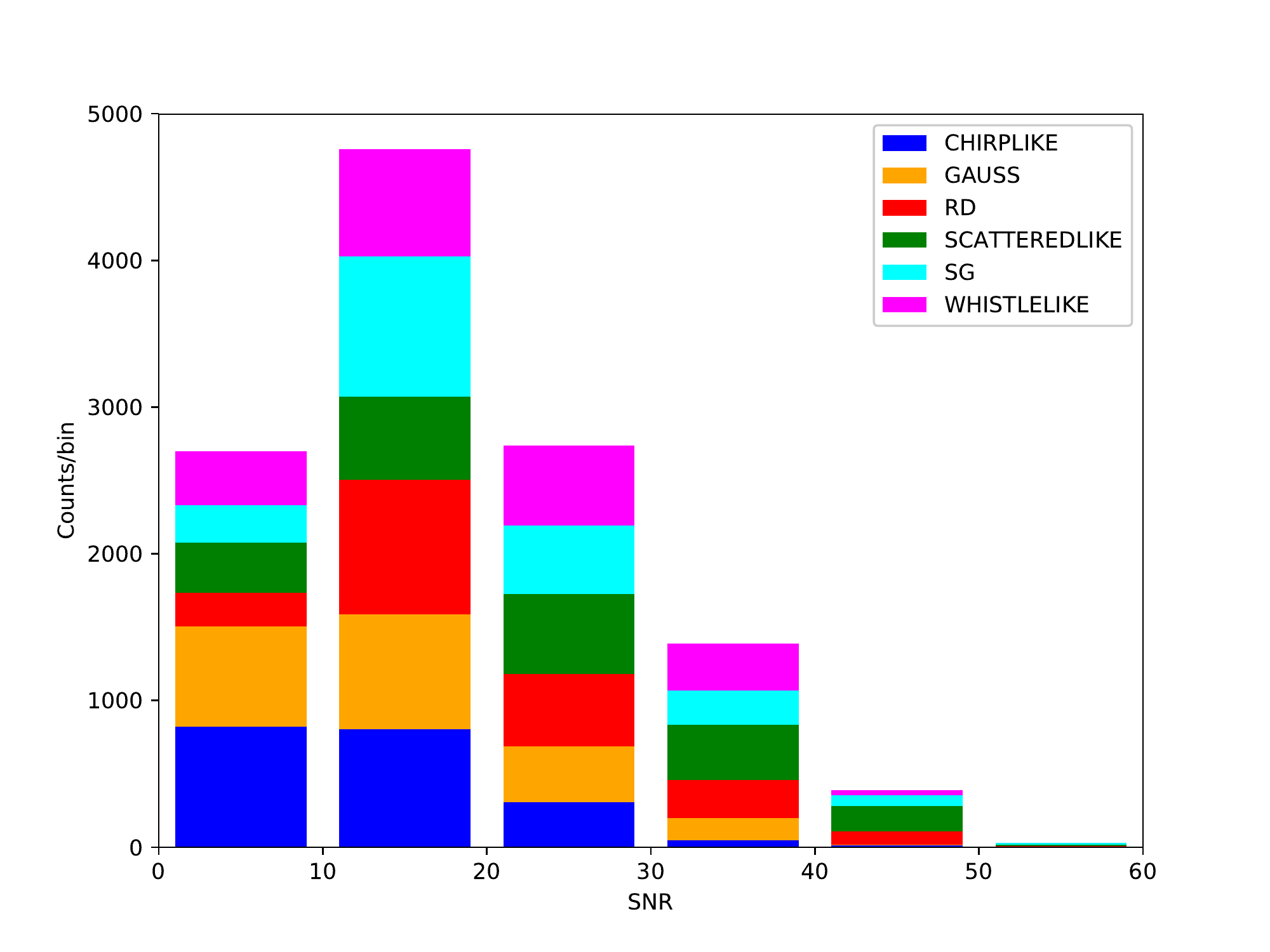}
\caption{\label{fig:snrdist}Distribution of SNR for the simulated glitch set.}
\end{figure}

Using the glitch families described in the previous section we have built a simulated set to test the classification pipeline. Glitches are added to the simulation of Gaussian noise derived from the sensitivity curve of the LIGO H1 detector.
The times t$_{0}$ at which the glitches occur are drawn from a Poisson distribution with mean rate of 0.5 Hz. With this choice it might be that two glitches are nearby in time, thus offering the opportunity to test the classification performance and robustness against nearby glitches.\\ 
The sampling rate of the simulated time series is 8192 Hz, which is half the sampling rate used for LIGO and Virgo \cite{2016transientnoise}. The original 16 kHz sampling rate could be used but will produce larger images and a large computational cost. We found that this choice is a good compromise between memory resources and the capability of highlighting fine details in glitches, that can be used for the classification.  Without loss of generality, when processing real data we can simply downsample the measured h(t) to the desired sampling rate.\\
The simulated glitch set has the purpose of testing the classification for glitches of different shapes in a wide range of duration, frequencies, and \ereviewme{signal-to-noise ratios (SNR)}. We therefore included in this set 2000 glitches for each family described in section \ref{sec:glitchfam}, \ereviewme{plus an additional 2000 images from the NOISE class (see sec.  \ref{sec:detection}) which adds up to} a total of 14000 samples. The summary of the parameters used for the simulations are reported in Table \ref{tab:set2}.\\
\begin{table}[ht]
\centering
\footnotesize
\begin{tabular}{@{}lll}
\br
Glitch class&Parameter&Value \\
\mr
\mr
Gauss (GAUSS)&$\tau$&4$\times$10$^{-4}$ -- 4$\times$10$^{-3}$ s\\
 &\ireviewme{hrss}&2$\times$10$^{-23}$ -- 1.5$\times$10$^{-22}$ Hz$^{-0.5}$\\
\mr
sine-Gaussian (SG)&f$_{0}$&50 -- 1500 Hz\\
  &Q&2 -- 20\\
&\ireviewme{hrss}&5$\times$10$^{-23}$ -- 1.5$\times$10$^{-22}$ Hz$^{-0.5}$\\
\mr
Chirp-like (CL)&M$_{1}$&1.4 -- 30 M$_{sol}$\\
 &M$_{2}$&1.4 -- 30 M$_{sol}$\\
&\ireviewme{hrss}&7$\times$10$^{-23}$ -- 7$\times$10$^{-22}$ Hz$^{-0.5}$\\
\mr
Ringdown (RD)&f$_{0}$&50 -- 1500 Hz\\
  &Q&2 -- 20\\
&\ireviewme{hrss}&5$\times$10$^{-23}$ -- 1.5$\times$10$^{-22}$ Hz$^{-0.5}$\\
\mr
Scattered Light-like (SL)&f$_{0}$&32 -- 64 Hz\\
  &$\tau$&0.2 -- 0.5 s \\
&\ireviewme{hrss}&2$\times$10$^{-23}$ -- 2$\times$10$^{-22}$ Hz$^{-0.5}$\\
\mr
Whistle-like (WL)&f$_{0}$&500 -- 3200 Hz\\
  &$\tau$&0.1 -- 0.3 s\\
&\ireviewme{hrss}&5$\times$10$^{-23}$ -- 3$\times$10$^{-21}$ Hz$^{-0.5}$\\
\br
\end{tabular}\\
\caption{\label{tab:set2}Summary of the parameters used to prepare the simulated glitch set}
\end{table}
In order to evaluate the accuracy of the algorithm as a function of the intensity of the transients, for each glitch we computed the optimal \ereviewme{SNR} based on matched filtering. \ireviewme{The values of hrss have been chosen in order to have a maximum SNR of $\sim$60.} The distribution of the SNR of the simulated glitches is shown in figure \ref{fig:snrdist}.
\section{Results}\label{sec:results}
\subsection{Image preparation and processing}
As anticipated in section \ref{sec:whitening}, in order to build our training \ereviewme{, validation,} and test sample we have applied a whitening procedure to the simulated time series and then we have built the image spectrograms over a time window of 2 seconds centered on the glitch time. The size of the output images is 241 pixel along the time axis and 513 pixels along the frequency axis. We then applied clipping and contrast stretching as described in section \ref{sec:preparation}.\\
In order to test the capability of the CNN to discriminate glitches from noise, we selected 2000 random windows of 2 seconds with no glitches and we built a seventh class, called NOISE, \ereviewme{that we described in section \ref{sec:detection} and added to our training, validation and test samples}.
\begin{figure}[ht]
\centering 
\subfloat[]{\includegraphics[width=0.45\textwidth]{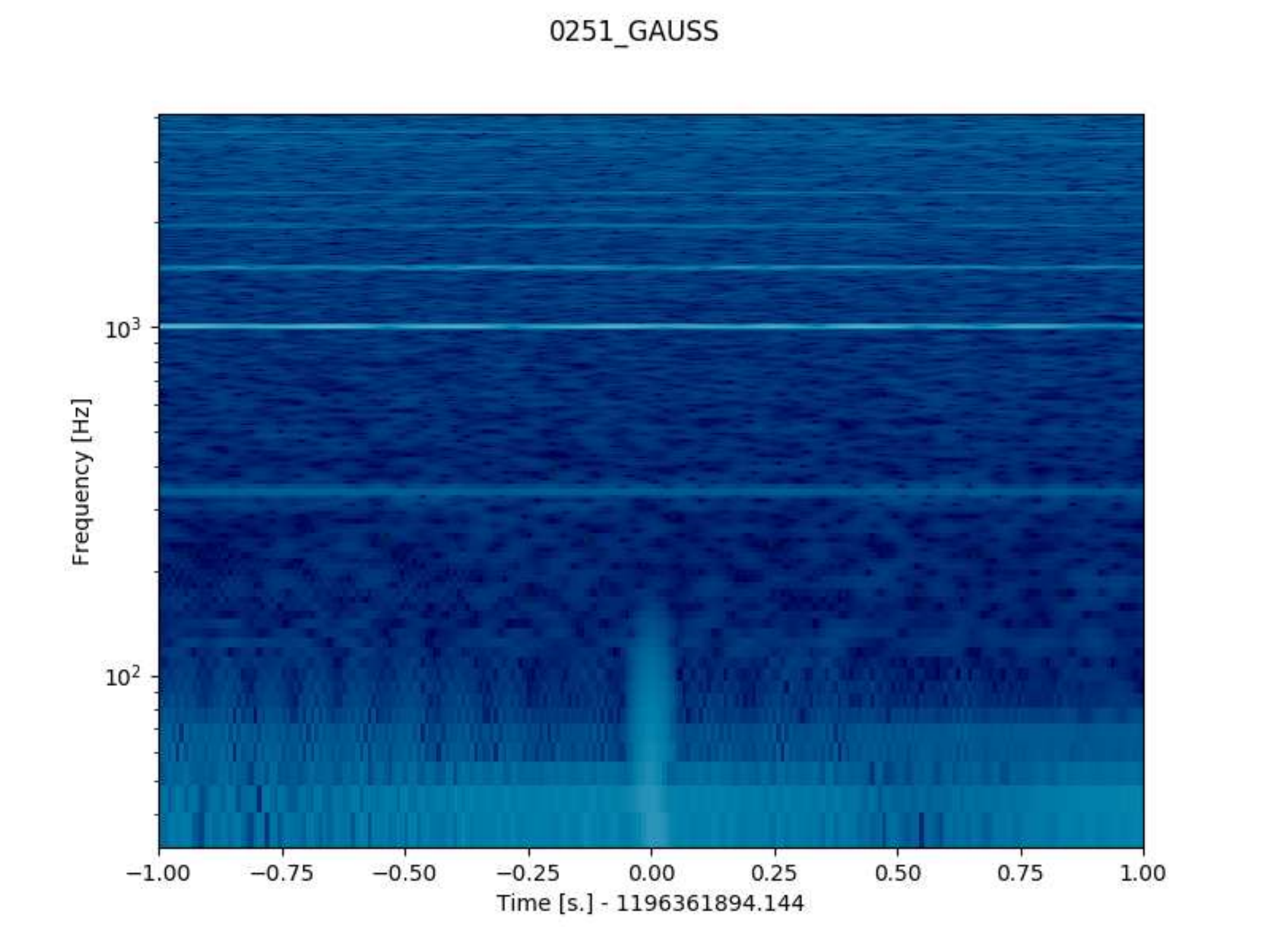}}
\subfloat[]{\includegraphics[width=0.45\textwidth]{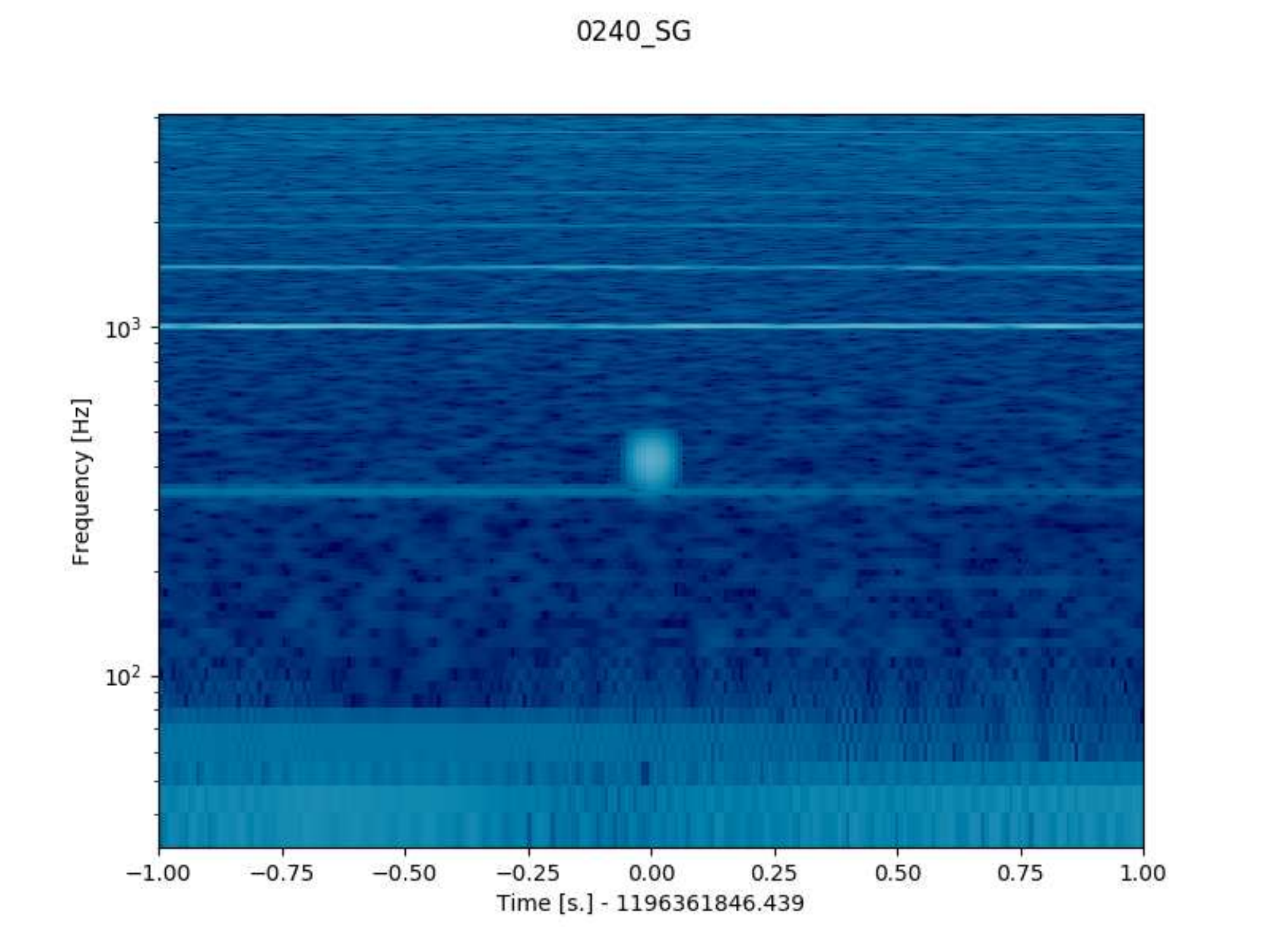}}
\hspace{0mm}
\subfloat[]{\includegraphics[width=0.45\textwidth]{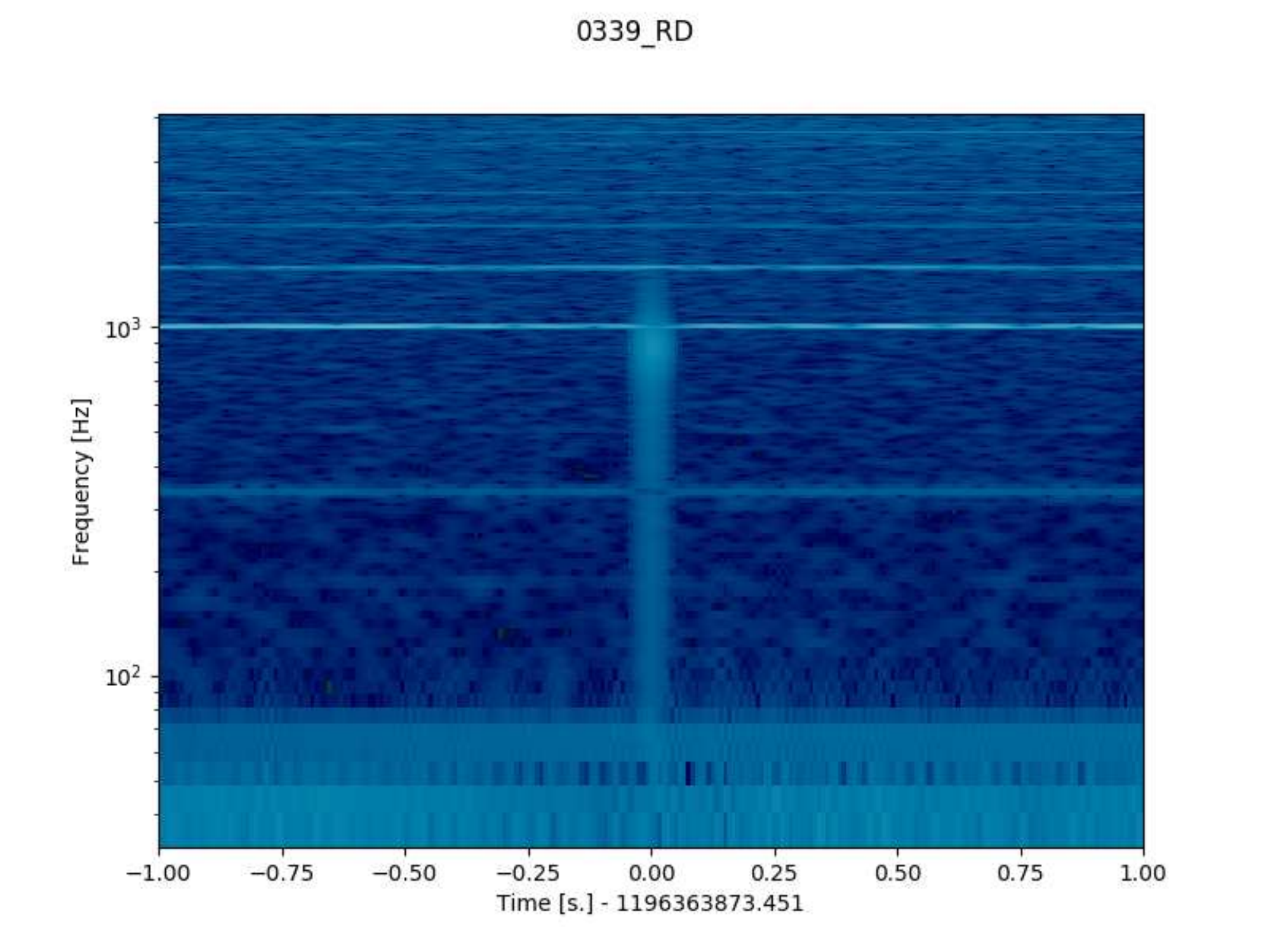}}
\subfloat[]{\includegraphics[width=0.45\textwidth]{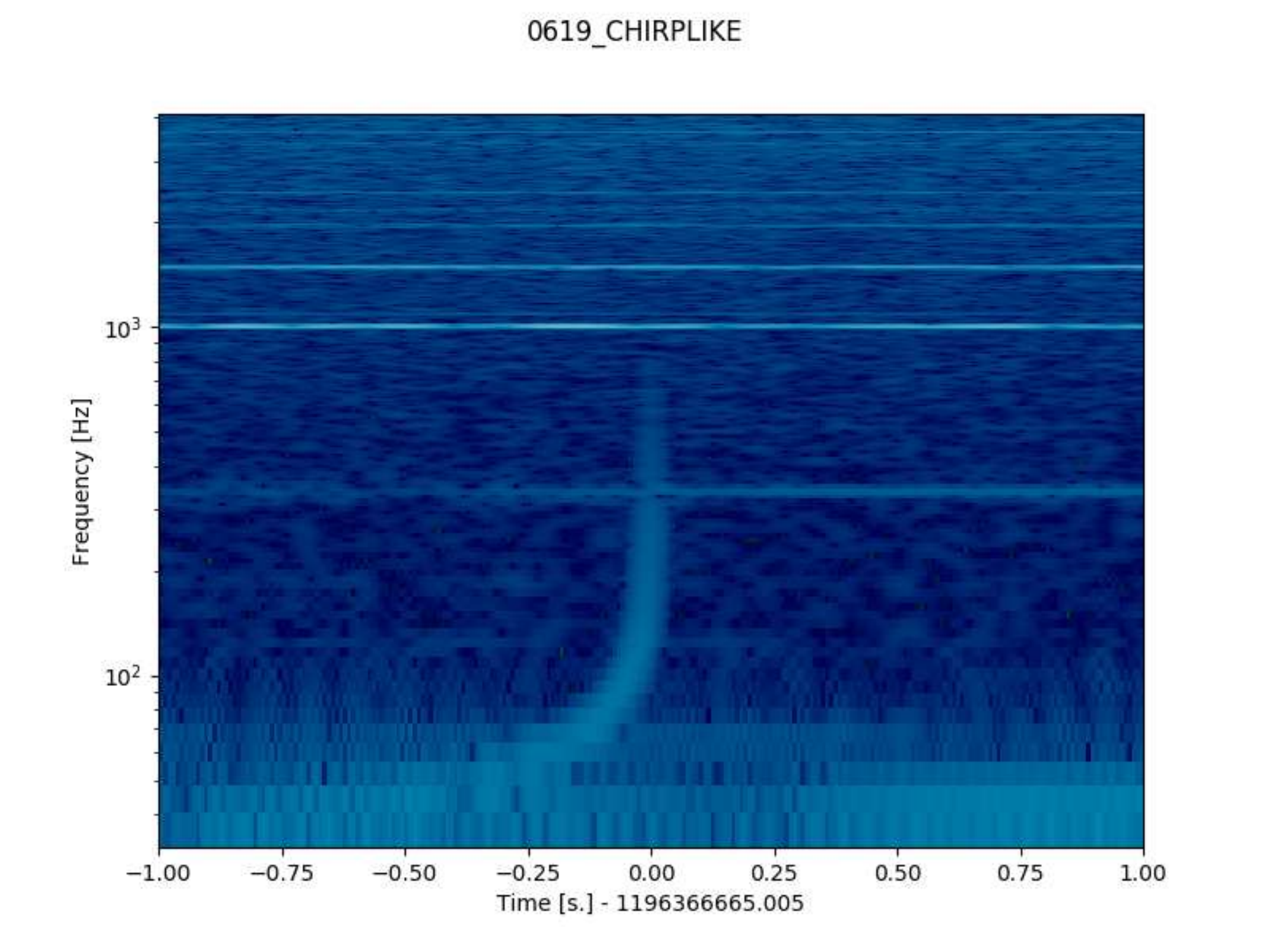}}
\hspace{0mm}
\subfloat[]{\includegraphics[width=0.45\textwidth]{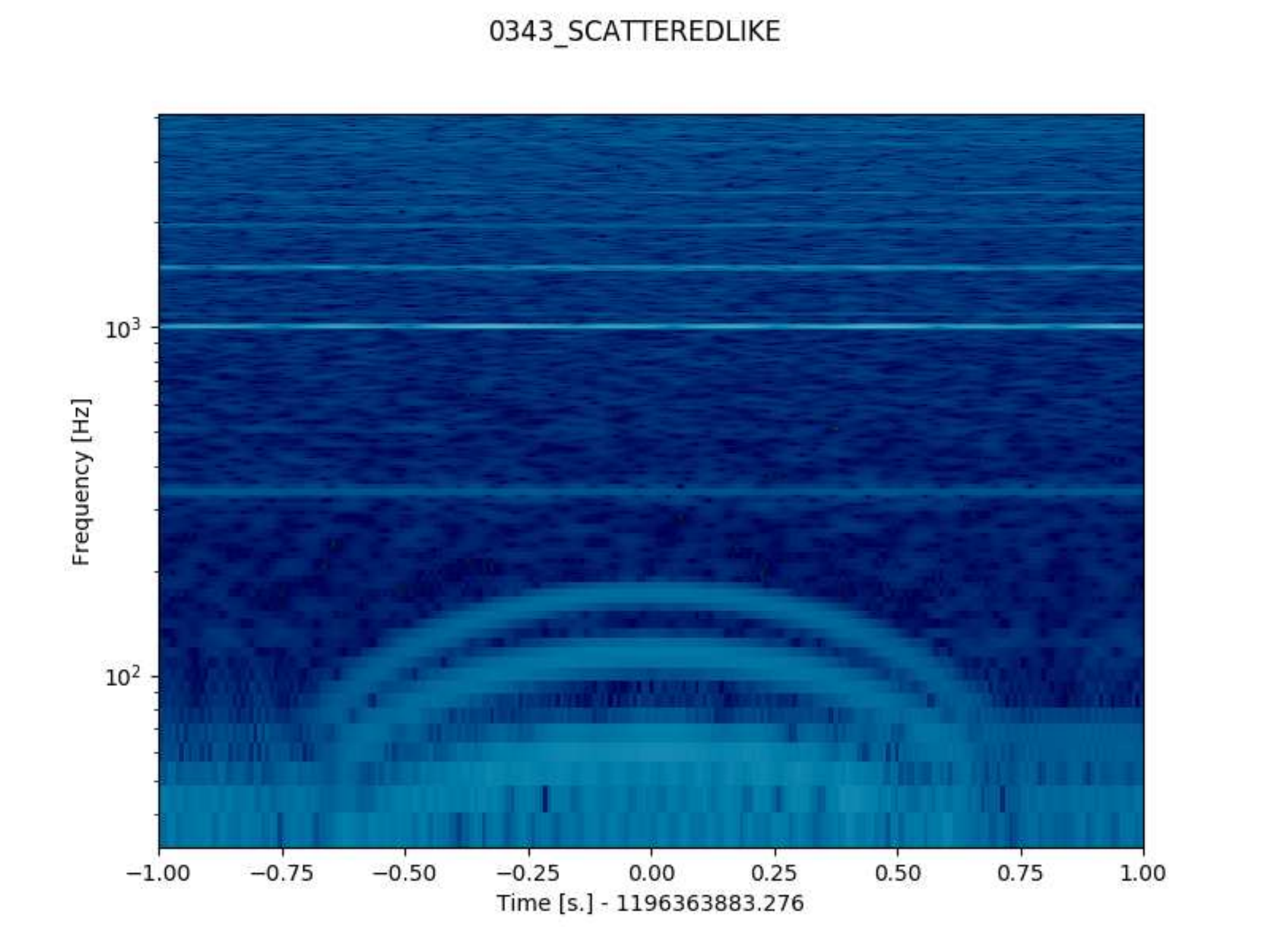}}
\subfloat[]{\includegraphics[width=0.45\textwidth]{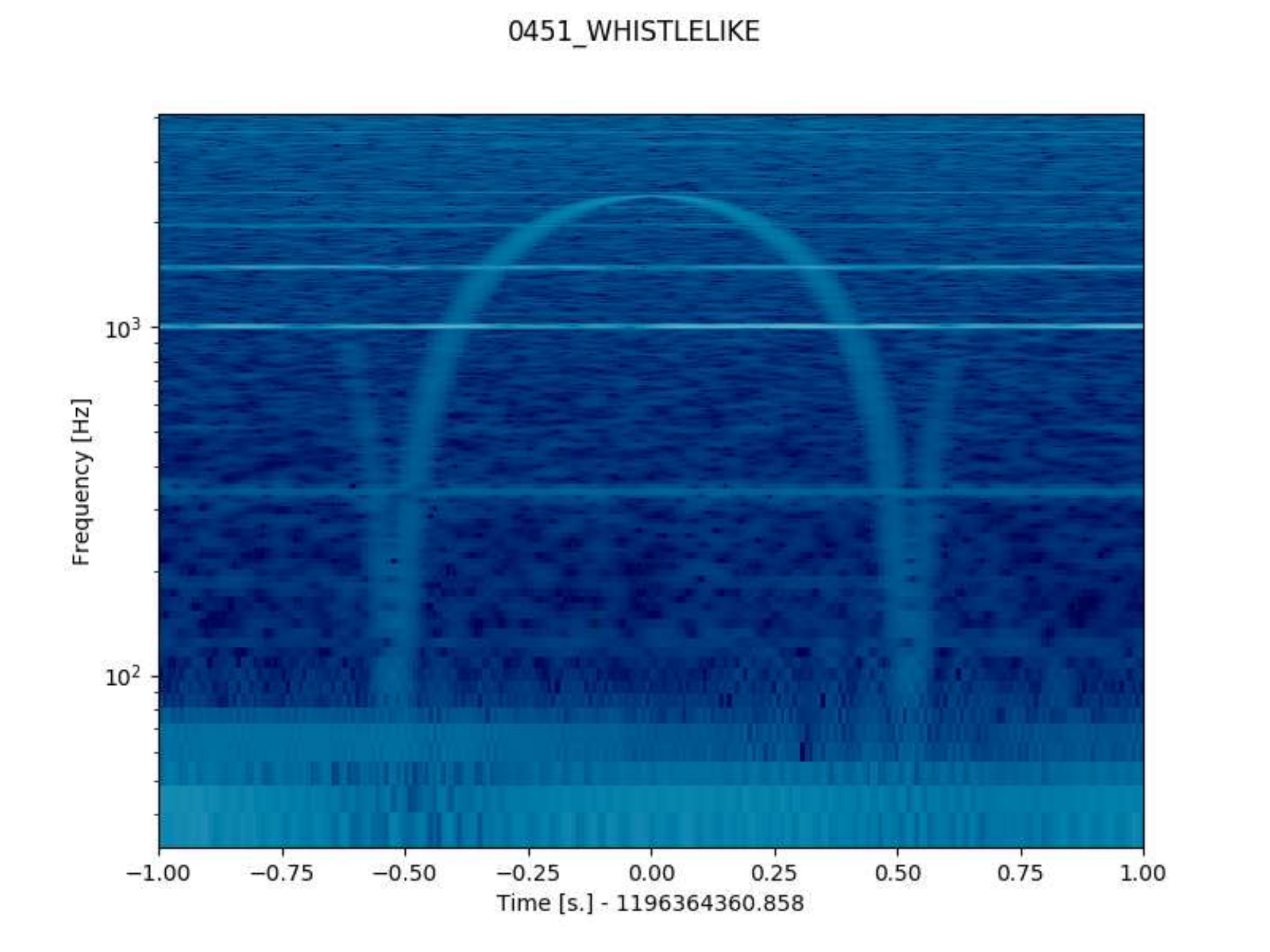}}
\caption{\label{fig:specgallery}Gallery of sample spectrograms obtained using simulated glitches from simulated sets 1 and 2. (a): \ireviewme{Gaussian \ereviewmetwo{(SNR$\sim$7) (b): Sine Gaussian (SNR$\sim$22)} (c) Ringdown (SNR=12) (d) Chirp-like (\ereviewmetwo{$\sim$8}) (e) Scattered light-like (SNR$\sim$24) (f) Whistle-like \ereviewmetwo{(SNR$\sim$10)}}. No whitening or image processing are applied.}
\end{figure}
We also scaled the sizes of images by a factor of \ereviewmetwo{0.55} on both axis. This scaling 
reduces the memory required without impacting significantly on the performance of the classification.\\
In order to show the impact of the whitening and image processing in cleaning the images, in Figure \ref{fig:specgallery} we show a gallery of spectrograms of simulated glitches without any processing. \ereviewmetwo{The images in this figure have been handpicked in order to better show the time-frequency charactestic evolution of each glitch family.}\\
Furthermore, Figure \ref{fig:specgalleryclean} shows the spectrograms of the same glitches of Figure \ref{fig:specgallery} after whitening and image processing steps have been applied. \ereviewmetwo{Although the whitening step removes most of the background noise, some residuals could be visibile at some frequencies, but they do not impact the classification task.}
\begin{figure}[ht]
\centering 
\subfloat[]{\includegraphics[width=0.45\textwidth]{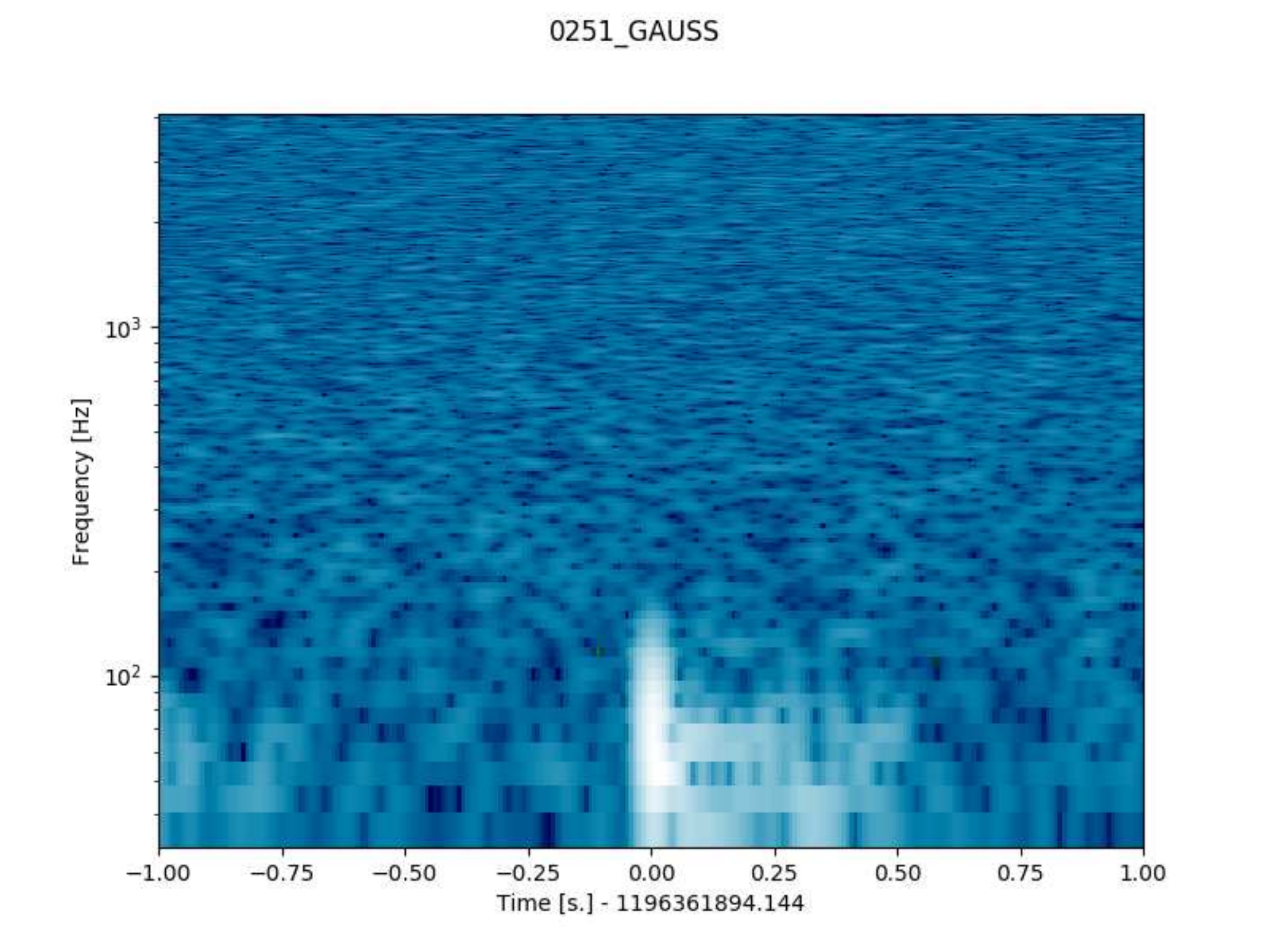}}
\subfloat[]{\includegraphics[width=0.45\textwidth]{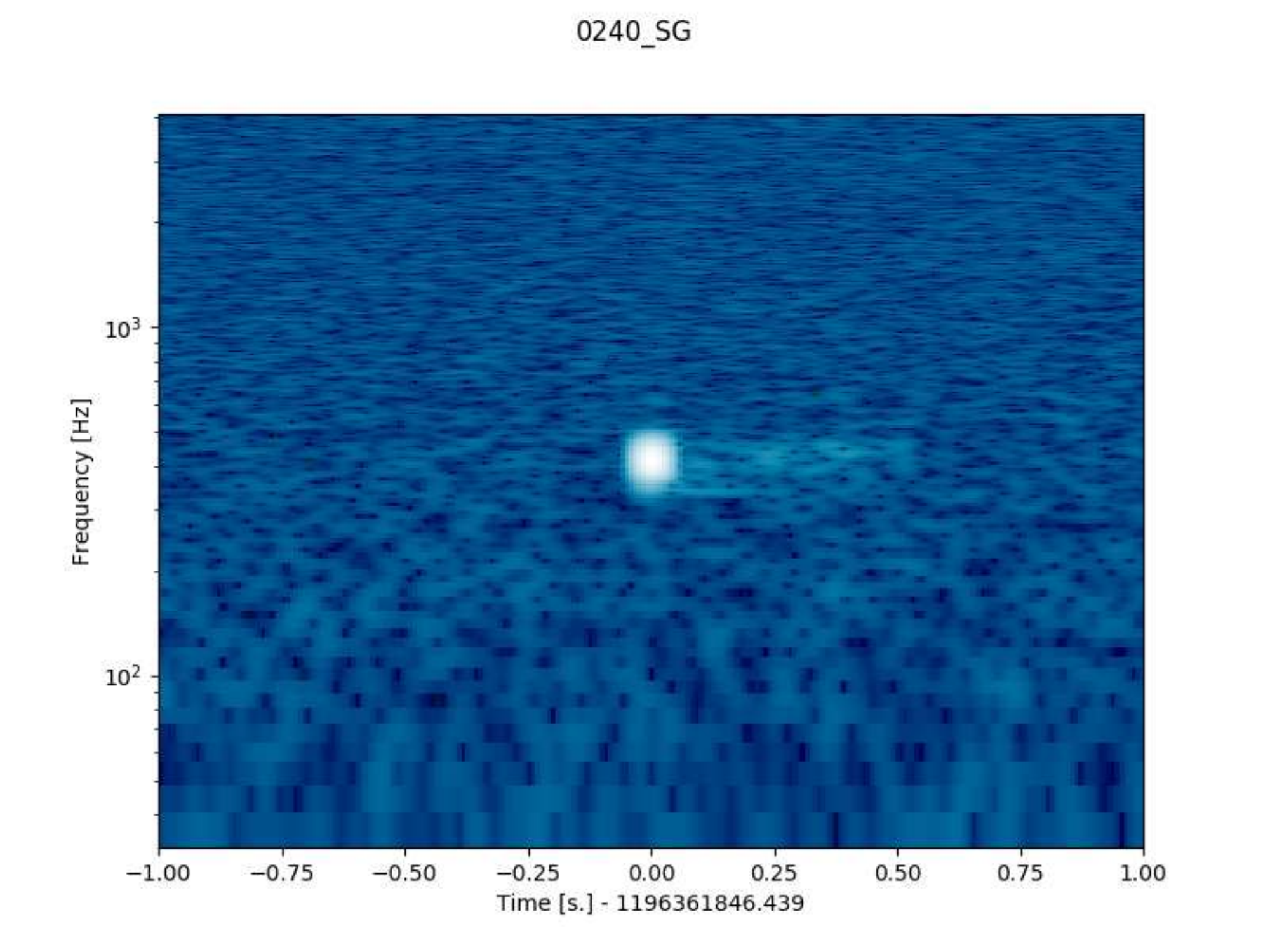}}
\hspace{0mm}
\subfloat[]{\includegraphics[width=0.45\textwidth]{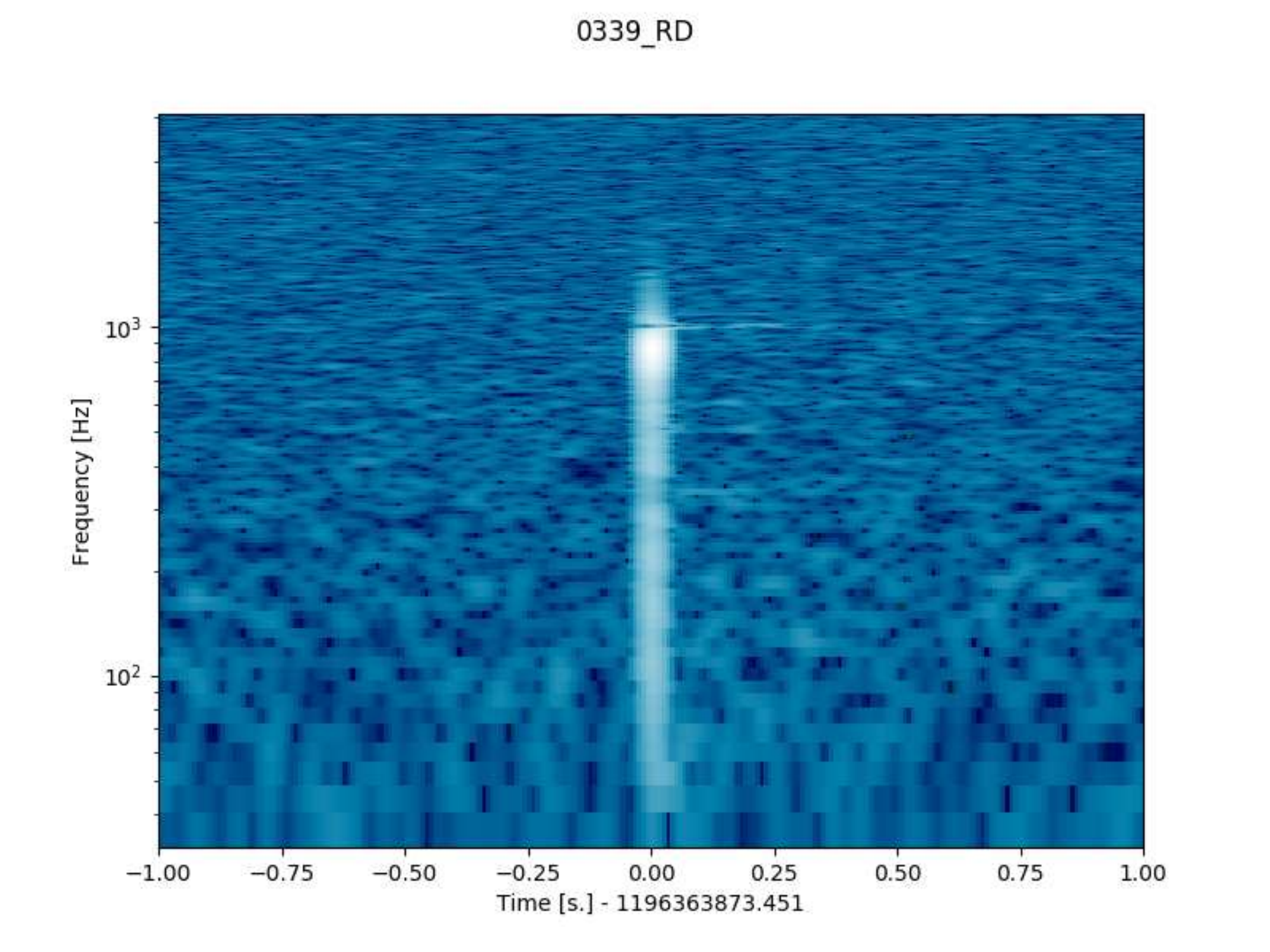}}
\subfloat[]{\includegraphics[width=0.45\textwidth]{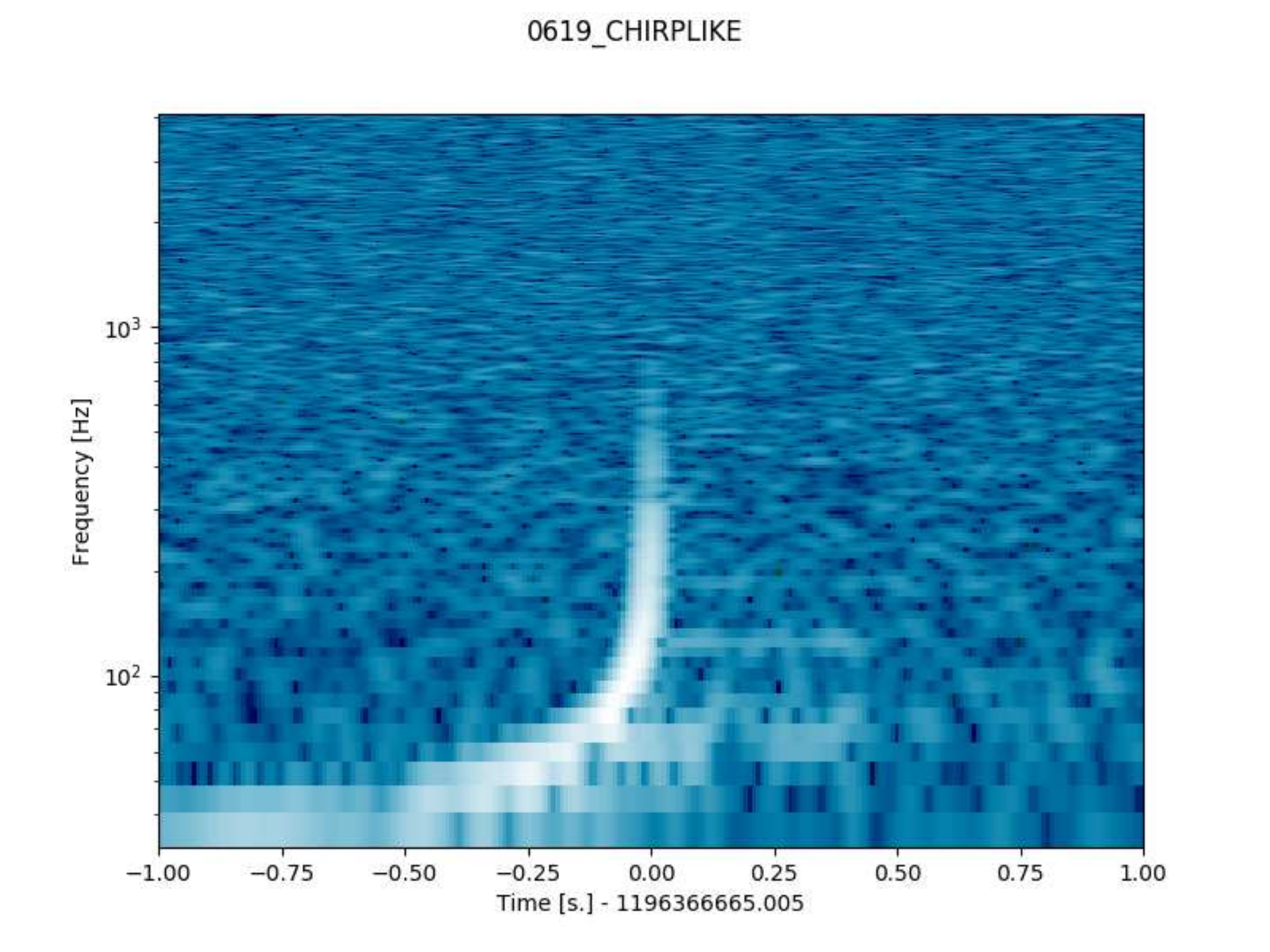}}
\hspace{0mm}
\subfloat[]{\includegraphics[width=0.45\textwidth]{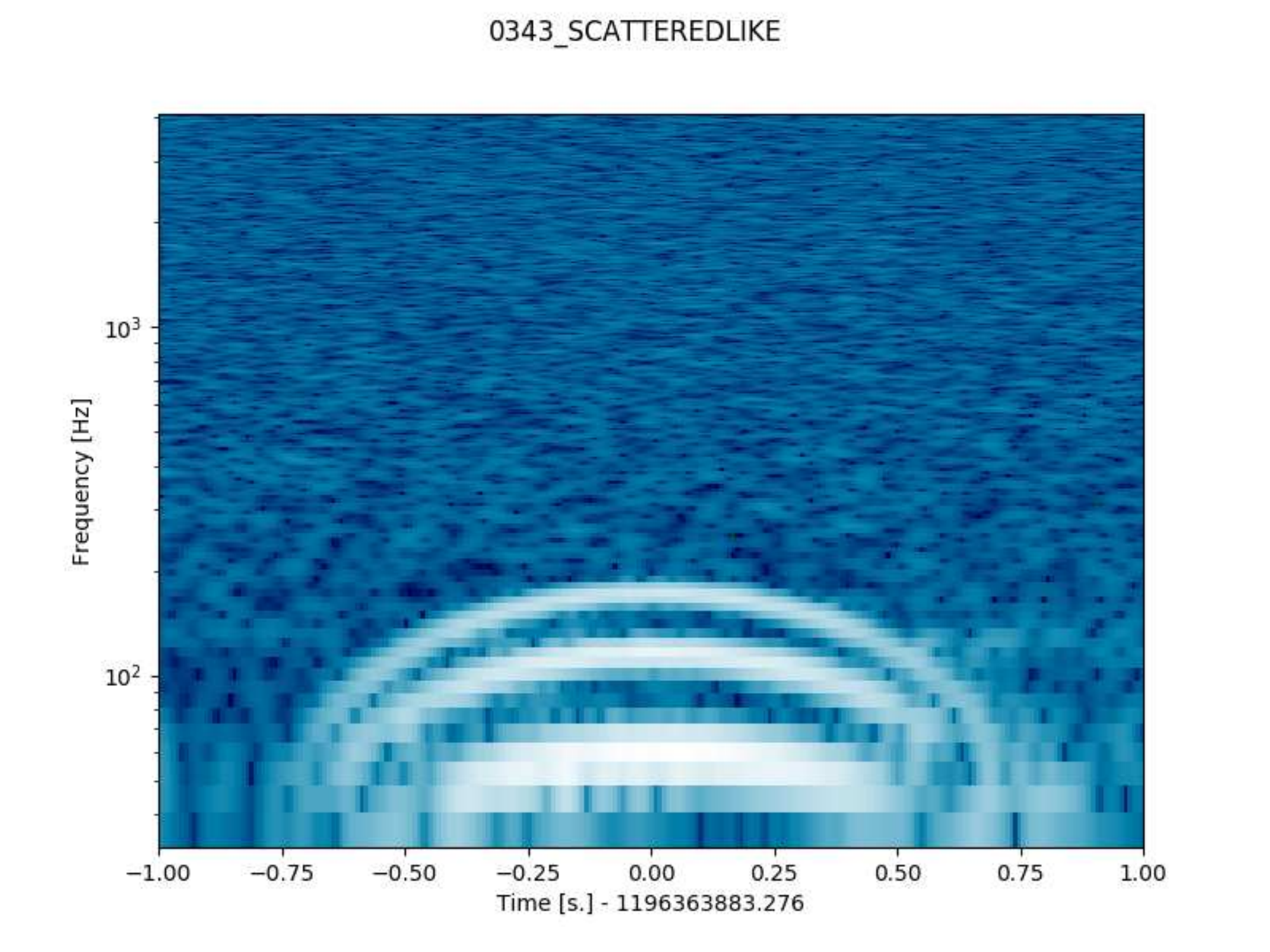}}
\subfloat[]{\includegraphics[width=0.45\textwidth]{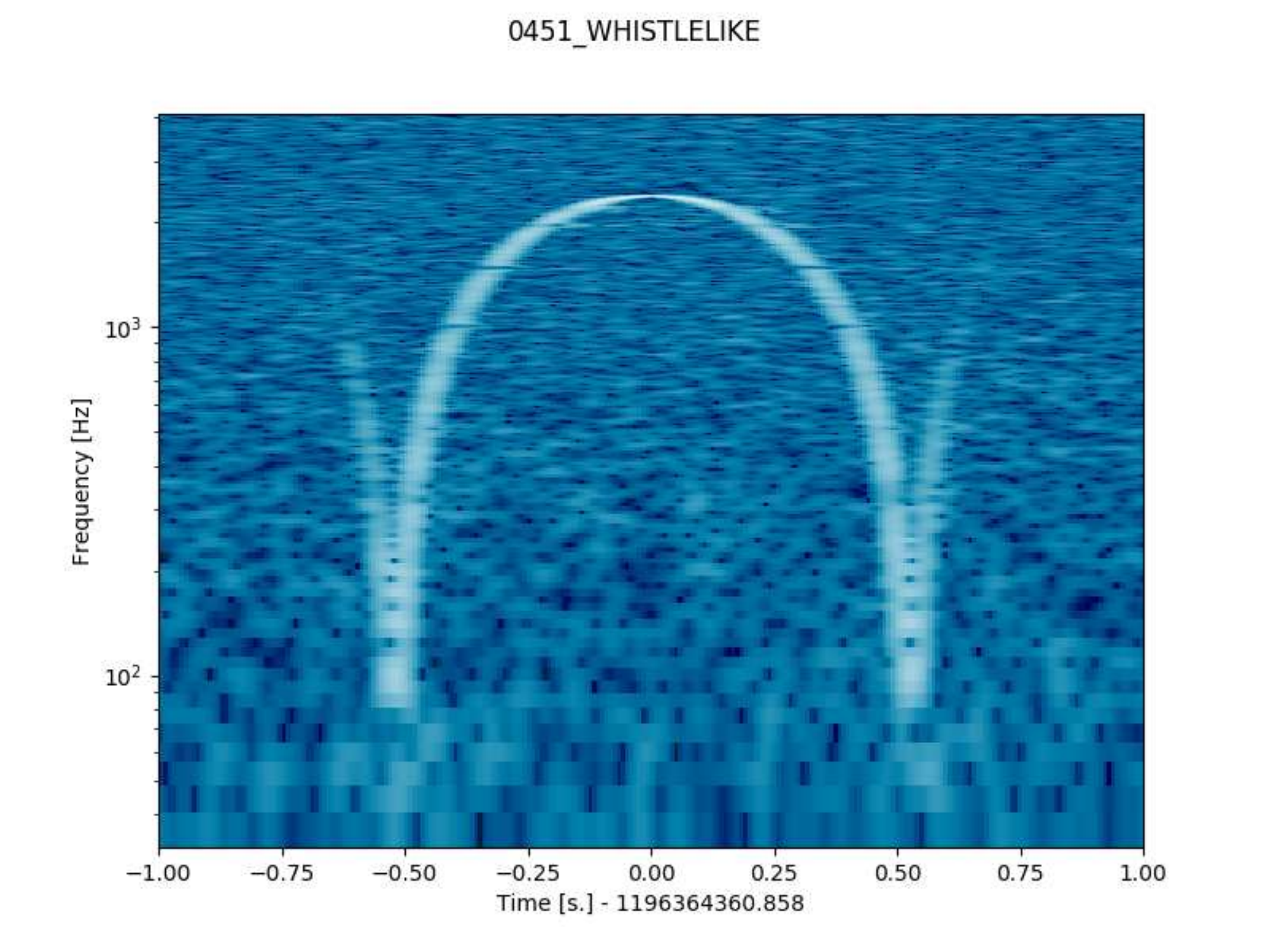}}
\caption{\label{fig:specgalleryclean}These spectrograms refer to the same glitches of Figure \ref{fig:specgallery}, but here the time series have been whitened and the images are been further cleaned as explained in section \ref{sec:whitening}. 
}
\end{figure}
\subsection{Detection and Classification}
\ereviewme{We used the 15\% of images as a test sample, 70\% as training one and 15\% as validation set. The training, validation and train set are stratified randomly selected.}\\ 
As described in section \ref{sec:implementaion}, the output of the CNN is a vector of probabilities of glitches belonging to each of the 7 classes (6 glitch families and NOISE). In order to assign the class to each glitch, we take the class with the highest probability. This choice can be modified in order to have hints on glitch belonging to new, unknown, classes, e.g. by assigning the class only if the probability is above a certain threshold.\\
\begin{table}[ht]
\begin{tabular}{@{}lccccc}
\br
Dropout strength&Accuracy\\
\mr
\mr
0.1&0.9983\\
0.2&0.9987\\
0.3&0.9987\\
0.4&0.9986\\
0.5&0.9982\\
0.6&0.9947\\
0.7&0.9799\\
0.8&0.7594\\
\mr
\end{tabular}
\caption{\label{tab:dropout}The table shows how the accuracy of the validation dataset depends on the the dropout strength. The scan in dropout strenght has been performed on the simulated glitch dataset used for this work.}
\end{table}
We have performed a tuning of the hyperparameters by studying the classification accuracy on the validation set. As an example, \ereviewmethree{Table \ref{tab:dropout}} shows a scan in dropout strength, in order to find the values giving the highest accuracy. From this plot it can be seen that the maximum is between 0.2 and 0.3, then we decided to use 0.25 as our dropout strength for the deep network of \ref{fig:architecture}. We also verified that for this value the accuracy still reaches the highest accuracy.\\
\ereviewmethree{During the hyperparameter tuning we tested different batch sizes of 32, 64 and 128, and selected the value of 128 since it gives the lowest fluctuations in the training curve. The accuracy has been computed using the validation set, and the optimal epoch has been chosen by looking at the evolution of the learning curve on the validation dataset. We have noticed that the accuracy is very high and having a further fine-tuning of the hyperparameters did not improve significantly more the accuracy.}

We then studied how the architecture described in section \ref{sec:implementaion} performed in comparison with simpler architectures. In particular, we considered a baseline machine learning algorithm (Support Vector Machine) and two simpler neural networks. The size of images are limited by the memory of the GPU card, therefore for the classification using SVM the images have been reduced \ereviewmethree{by a factor of 0.55, the same used for the CNNs.} \ereviewmethree{We used a linear SVM model and adopted a L2 regularization term in the cost function. We performed a search over the penalty parameter C and found 1.0 as the optimal value that maximizes the validation accuracy, estimated using a 10-fold cross validation procedure.}\\
We then performed a comparison with a shallow network consisting of a single, 32-deep CNN hidden Keras Conv2D layer, followed by two fully connected Keras Dense layers with 64 and N$_{c}$+1 neurons respectively, where $N_{c}$ is the number of glitch classes and the additional one is inserted to represent the NOISE class. We also added to this comparison a more complex, two-layers network resembling the CNN blocks of our deep configuration described in section \ref{sec:implementaion}, made of two 16-deep Keras Conv2D layers followed by a Keras MaxPooling2D, a Keras Dropout layer. The block is attached to two Keras Dense layers with 256 and $N_{c}$+1 neurons respectively.\\
Table \ref{tab:comparison} reports the overall classification metrics for our deep network compared with other configurations, where it can be seen that a deeper network is performing better. In fact, although the baseline SVM algorithm is performing well on glitches that exhibit very different time-frequency behaviour, it has some difficulties in comparing glitches with similar shapes, such as GAUSS, SG and RD.\\
\begin{table}[ht]
\begin{tabular}{@{}lccccc}
\br
Metric&Accuracy&Precision&Recall&F1 score&Log loss\\
\mr
\mr
\ereviewmetwo{SVM}&0.971&0.972&0.971&0.971&0.08\\
\mr
Shallow CNN &0.986&0.986&0.986&0.986&0.04\\
\mr
1 CNN block&0.991&0.991&0.991&0.991&0.02\\
\mr
\ereviewmetwo{3 CNN blocks}&0.998&0.998&0.998&0.998&0.008\\
\mr
\end{tabular}
\caption{\label{tab:comparison}\ereviewme{Comparison of performance between the selected deep architecture presented in section \ref{sec:implementaion} and simpler CNN architecture. The metrics are evaluated on the images belonging to the validation sample.}}
\end{table}

\begin{figure}[ht]
\centering 
\subfloat[]{\includegraphics[width=0.5\textwidth]{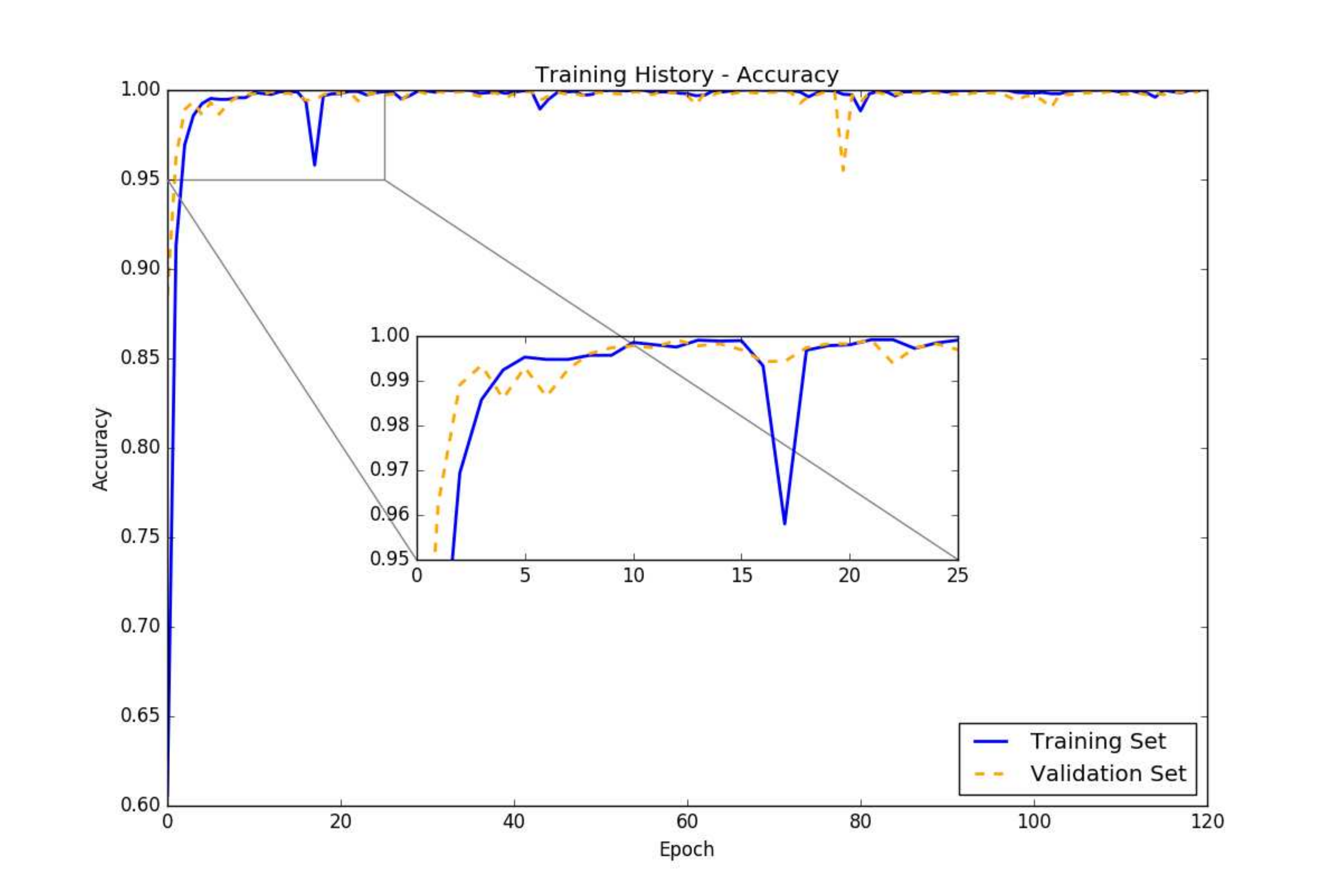}}
\subfloat[]{\includegraphics[width=0.5\textwidth]
{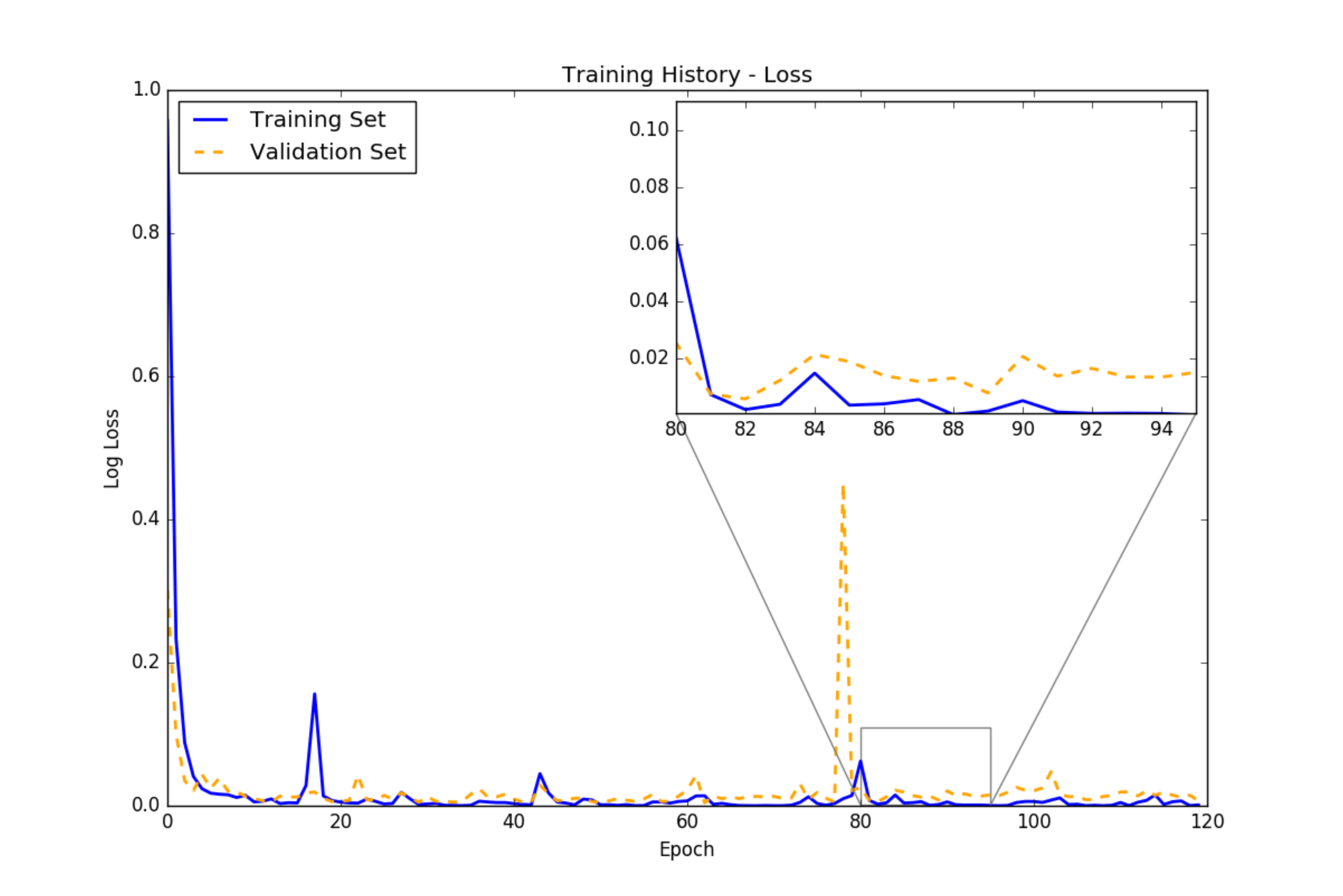}}
\caption{\label{fig:learningcurve}Evolution of accuracy (left) and log loss (right) as a function of the training epoch, applied to the training and validation sample. \ereviewmethree{The insets show a zoom on the learning curve, showing respectively the regions where the accuracy reaches a plateau and where the log loss of the validation sample start to increase significantly}}
\end{figure}

For the training phase of our 3-block network we used mini batches of \ereviewmetwo{64} images. Figure \ref{fig:learningcurve} shows the training history curves as a function of training epoch, evaluated on the training and validation samples. The accuracy rises fast with the training epoch and reaches a plateau after $\sim$ 20 iterations. The evolution of the log loss of the validation sample is rapidly decreasing and starts rising \ereviewmethree{significantly} \ereviewmetwo{from $\sim$80 iterations, indicating the beginning of an overfitting regime. We therefore decided to stop our training after 70 epochs.} The accuracy and log loss applied to the test sample is consistent with that obtained on the validation sample, showing that our network and its training procedure is reliable.
\ireviewme{Table \ref{tab:metrics} shows a summary of some important accuracy metrics used to evaluate machine learning classification algorithms, \ereviewme{applied on the test set containing 15\% of the data.}}
\begin{table}
\begin{tabular}{@{}lccccc}
\br
Metric&Accuracy&Precision&Recall&F1 score&Log loss\\
\mr
\mr
ALL&0.998&0.998&0.998&0.998&0.006\\
\mr
GAUSS&0.997&0.997&0.997&0.997&0.008\\
\mr
SG&0.997&0.997&0.997&0.997&0.01\\
\mr
CHIRPLIKE&1.0&1.0&1.0&1.0&2$\times10^{-6}$\\
\mr
RD&0.993&0.993&0.993&0.993&0.003\\
\mr
SCATTEREDLIKE&1.0&1.0&1.0&1.0&5$\times$10$^{-5}$\\
\mr
WHISTLELIKE&1.0&1.0&1.0&1.0&3$\times$10$^{-8}$\\
\mr
NOISE&1.0&1.0&1.0&1.0&5$\times$10$^{-7}$\\
\br
\end{tabular}\\
\caption{\label{tab:metrics}Summary of metrics that describe the performance of the algorithm for the classification of the simulated glitch set. \ereviewme{Each metric is evaluated on the test set}}
\end{table}

\begin{figure}[ht]
\centering 
\subfloat[]{\includegraphics[width=0.58\textwidth]{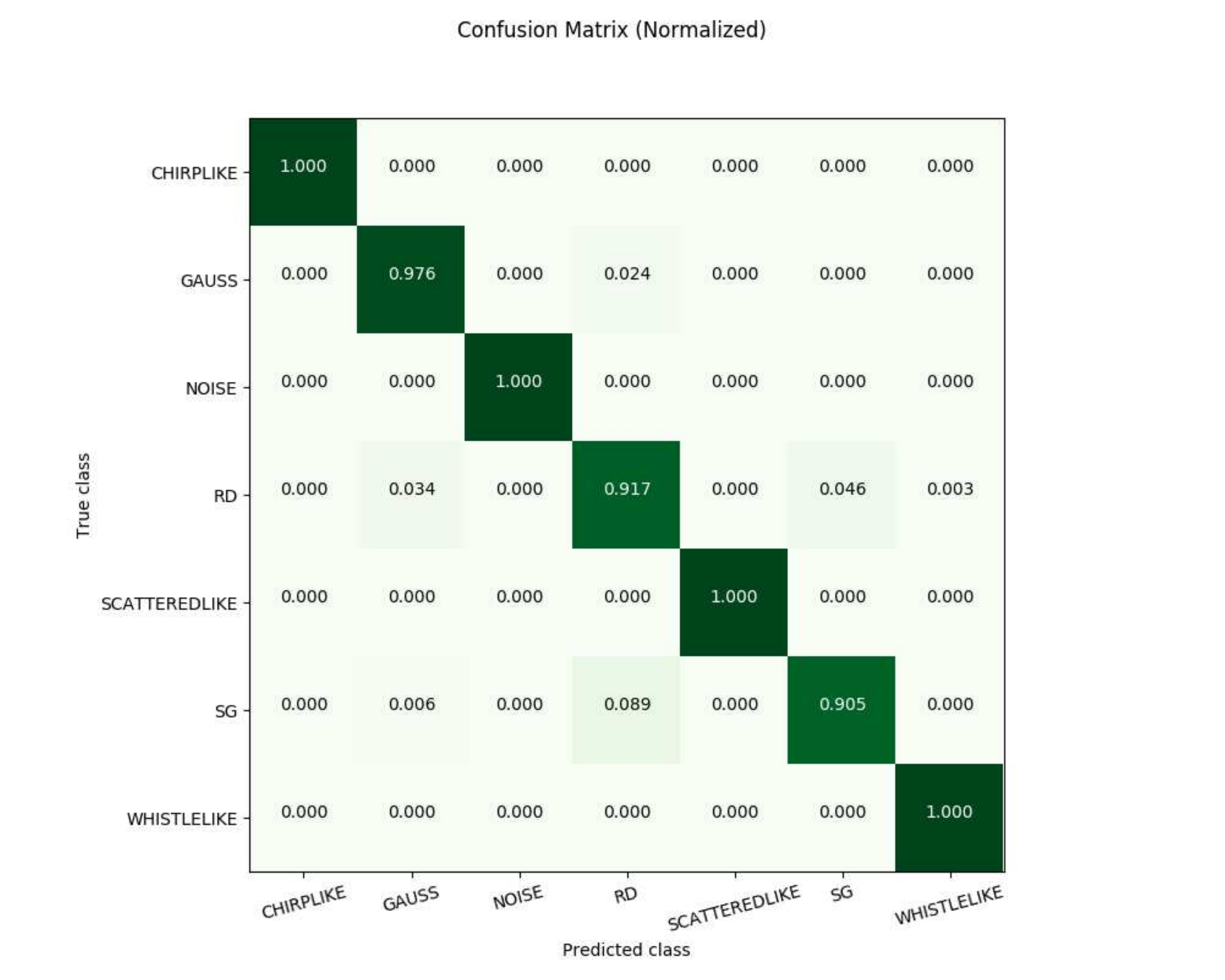}}
\subfloat[]{\includegraphics[width=0.58\textwidth]{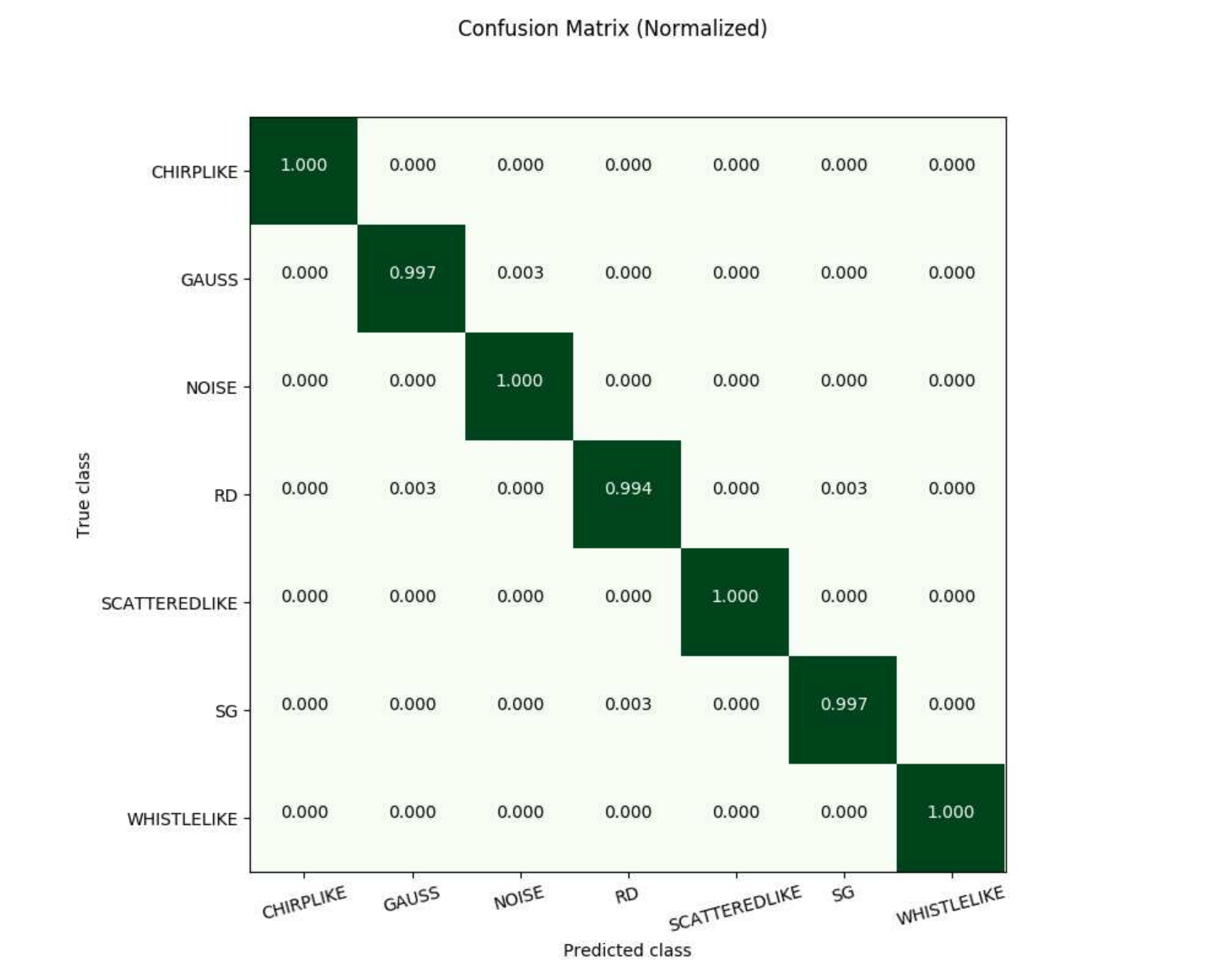}}
\caption{\label{fig:cm_a}Confusion matrix for the multi-label classification based on images, evaluated on the test set.\ereviewmetwo{Left: baseline SVM classification algorithm. Right: CNN-based deep network described in section \ref{sec:implementaion}}}
\end{figure}

\begin{figure}[ht]
\centering 
\subfloat[]{\includegraphics[width=0.6\textwidth]{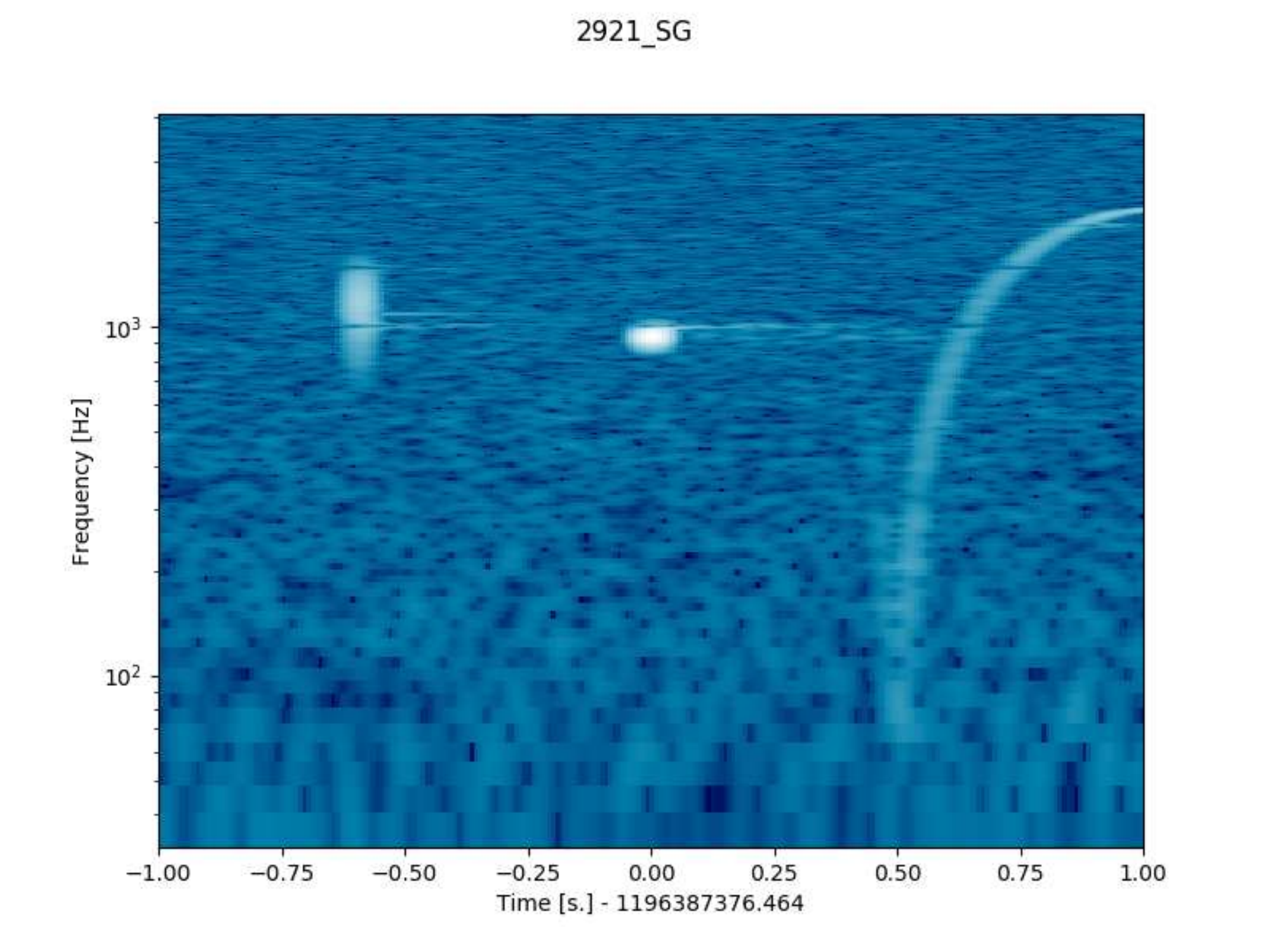}}
\caption{\label{fig:manyglitches} \ereviewmetwo{A more complex case of an image containing more than one glitch, namely a GAUSS, a SG and a WHISTLELIKE glitch.}}
\end{figure}

We then computed the confusion matrix for the classification, and the results are shown in Figure \ref{fig:cm_a} \ereviewmetwo{for both the SVM and the deep CNN classifier. We see that the accuracy on each class is greater than \ireviewme{99}\%. Both methods are excellent in distinguishing NOISE from glitches, but the CNN shows higher performance than SVM in distinguishing glitches with similar time-frequency behaviour such as GAUSS,SG and RD.}
Furthermore, it can happen that two glitches appear close in time and as a consequence their time-frequency evolutions will enter in the same image and make the classification a bit harder. However, the capability of CNNs to distinguishing features is making the algorithm robust against these cases. As an example, Figure \ref{fig:manyglitches} shows a RD, SG and WHISTLELIKE glitch in the same image, and the algorithm computes the correct class (SG) \ereviewmethree{, since the output probability of belonging to that class is 1.0, while that of the other classes is $\sim$ 0.}





\section{Discussion}\label{sec:discussion}
The results of the previous section show that image-based deep learning algorithms are very powerful in discriminating glitch transients from  noise in gravitational wave interferometers. Furthermore, the CNNs use the time-frequency dependence of the glitches to provide high accuracy in classification of glitches among different classes.\\ 
In order to reach high classification accuracy, in particular for low-SNR glitches, it is important to remove the contribution of the detector noise in the time series. Therefore, for our simulated glitches we applied a whitening procedure and an image processing stage, in order to enhance the visibility of the glitches in the images. In this way,  $>$\ireviewme{99\% accuracy can be achieved}.\\ 
Previous works have shown how CNN-based algorithms are performing better in classification tasks with respect to other machine learning algorithms such as support vector machines, logistic regression or random forest \cite{2017george}. Furthermore, the capability of CNNs to extract features at various scales makes them more efficient with respect to the standard, fully-connected neural networks. This capability of extracting features in the images and use them is enhanced by some basic image processing steps.\\
In particular, we found that the capability of CNNs to look for features in the images has an important impact on the classification of time intervals containing glitches close to each other. Even in this case, the CNN was able to correctly recognize the glitch in the center of the image.\\
The duration of the training phase depends on the number of images in the training set, \ereviewme{but this phase} can be run on a dedicated GPU card, thus optimizing the resources. In our cases, the training of the architecture of section \ref{sec:implementaion} over \ereviewmetwo{70 epochs required about $\sim$5 hours} on a desktop PC equipped with a NVIDIA GeForce GTX 780 GPU card. Once the training is completed, the algorithm can classify each image within few microseconds. This approach is thus particularly important for low-latency applications, such as real-time, online recognition of glitches.\\
The parameters and weights of the neural network computed during the training phase can be stored on disk and loaded when necessary. For the dataset of 14000 simulated glitches, the size of the model is $\sim$ 50 Mb. This is quite small, compared to the size of the spectrograms used ($\sim$13 Gb) and the spectrogram images ($\sim$ 2.4 Gb).
\section{Conclusions}\label{sec:conclusions}
In this paper we have presented a new algorithm for the classification of transient noise glitches in advanced gravitational wave interferometers. The detection and identification of glitches are important tasks for the characterization of these complex detectors, in order to understand the origin of their noise sources and improve the sensitivity.\\
The algorithm has been built using convolutional neural networks, a class of deep learning algorithms very efficient for image processing and recognition. We have tested the proposed architecture against a set of simulated glitches superimposed to a colored Gaussian noise that reproduces to a good deal the actual noise in the advanced gravitational wave interferometers.\\
The presented results are encouraging and show the large potential of these algorithms for detector characterization. Without any particular need of optimization, the results show a $>$\ireviewme{99}\% accuracy on the glitches simulated over a larger parameter space.\\
As noted also in previous works, this approach largely benefits from fast hardware available today, such as fast CPU and GPUs, as well as software developed for deep learning applications. This technology is intensively used in many fields and now is getting more and more important also in scientific research. In fact, the requirements in term of memory or disk space is within the reach of commercial hardware available today, making these algorithms very promising for various field of research.\\
Thanks to its flexibility, CNNs can be easily extended, e.g. to looking at new glitch families occurring in the detector and provide an automatic classification of them.\\
The goal of the algorithm is to provide an optimal automatic classification of glitches, but it could be customized to look for faint signals, e.g. from astrophysical sources \cite{2017george}.
The presented algorithm is able to process glitches in an autonomous way but can be combined with external pipelines for the detection of transient signals. We plan to run this algorithm on the real data collected by the Advanced LIGO and Advanced Virgo interferometers during observation runs. This CNN-based method is therefore very promising as a new, powerful tool to investigate and characterize the transient noise in the era of advanced gravitational wave detectors.
\section{Acknowledgements}
We would like to thank the machine learning group and the detector characterization groups of the LIGO Scientific Collaboration and the Virgo Collaboration for the useful discussions of this work. We gratefully acknowledge the support of NVIDIA Corporation with the donation of the GeForce GTX 780 GPU used for this research. \ereviewme{We also thank Marco Cavagli\`a for the internal LIGO-Virgo review and the anonymous referees for the insightful and useful comments and suggestions}. This paper has been assigned the Virgo Document VIR-0669A-17 and LIGO Document P1700254.
\newpage
\printbibliography

\end{document}